%% file: rcs2327_ms.tex
\documentclass{emulateapj}

\usepackage{color}
\usepackage{url}

\newcommand{\hst}{{\it HST}}
\newcommand{\HST}{{\it HST}}
\newcommand{\RCS}{RCS2}

\newcommand{\OII}{[O II]$\lambda\lambda3727$}

\newcommand{\uber}{RCS2327}
\newcommand{\clustername}{RCS2327}
\newcommand{\clusterz}{0.6986}

\newcommand{\longuber}{RCS2~J232727.6-020437}
\newcommand{\elgordo}{ACT-CL J0102--4915}
\newcommand{\Msun}{M$_\sun$}
\newcommand{\msun}{M$_\sun$}
\newcommand{\etal}{{et~al.}~}
\newcommand{\rtfh}{\mbox{$R_{\mbox{\scriptsize 2500}}$}}

\newcommand{\h}{${ h}_{70}^{-1}$}

\newcommand{\Px}{$1.17 ^{+0.47}_{-0.24}$}
\newcommand{\Py}{$7.42 ^{+1.42}_{-0.63}$}
\newcommand{\Pe}{$0.26 ^{+0.04}_{-0.06}$}
\newcommand{\Ptheta}{$102.0 ^{+0.4}_{-1.0}$}
\newcommand{\Prs}{$486 ^{+ 62}_{-266}$}
\newcommand{\Palpha}{$1.44 ^{+0.09}_{-0.62}$}
\newcommand{\Pc}{$3.40 ^{+5.09}_{-0.40}$}

\newcommand{\MarcA}{$5.8 ^{+0.4}_{-0.2}$}
\newcommand{\MarcB}{$3.4 ^{+0.3}_{-0.1}$}
\newcommand{\MrfhX}{$6.2 ^{+2.1}_{-1.4}$}
\newcommand{\MarcAX}{$4.0 ^{+1.2}_{-0.8}$}
\newcommand{\MarcBX}{$2.0 ^{+0.5}_{-0.4}$}
\newcommand{\Mfhkpc}{$8.0 ^{+0.6}_{-0.4}$}

\newcommand{\MarcAm}{$5.8 ^{+0.4}_{-0.2}\times 10^{14}$\msun}
\newcommand{\MarcBm}{$3.4 ^{+0.3}_{-0.1}\times 10^{14}$\msun}

\newcommand{\Pasfit}{$r_s=575.4 -906.8\alpha +583.3\alpha^2$}
\newcommand{\Pacfit}{$c=16.1 -8.8\alpha$}

\slugcomment{ApJ in press: draft date \today}

\shorttitle{Mass Measurements of RCS2~J232727.6-020437}
\shortauthors{Sharon et al.}

\begin{document}

\title{A Multi-Wavelength Mass Analysis of RCS2~J232727.6-020437, a
  $\sim3 \times 10^{15}$\Msun ~Galaxy Cluster at z=0.7\footnote{Based on
    observations obtained with : MegaPrime/MegaCam, a joint project of
    CFHT and CEA/DAPNIA, at the Canada-France-Hawaii Telescope (CFHT)
    which is operated by the National Research Council (NRC) of
    Canada, the Institut National des Science de l'Univers of the
    Centre National de la Recherche Scientifique (CNRS) of France, and
    the University of Hawaii; the NASA/ESA Hubble Space Telescope,
    obtained from the data archive at the Space Telescope
    Institute. STScI is operated by the association of Universities
    for Research in Astronomy, Inc. under the NASA contract NAS
    5-2655; the 6.5\,m Magellan telescopes located at Las Campanas
    Observatory, Chile; }}

\author{K. Sharon\altaffilmark{1}}
\author{M.D. Gladders\altaffilmark{2,3}}
\author{D.P. Marrone\altaffilmark{4}}
\author{H. Hoekstra\altaffilmark{5}}
\author{E. Rasia\altaffilmark{6,7}}
\author{H. Bourdin\altaffilmark{8}}
\author{D. Gifford\altaffilmark{1}}
\author{A.K. Hicks\altaffilmark{9}}
\author{C. Greer\altaffilmark{4}}
\author{T. Mroczkowski\altaffilmark{16,17}}%
\author{L.F. Barrientos\altaffilmark{10}}
\author{M. Bayliss\altaffilmark{11,12}}
\author{J.E. Carlstrom\altaffilmark{2,3}}
\author{D.G. Gilbank\altaffilmark{13}}
\author{M. Gralla\altaffilmark{12,14}}
\author{J. Hlavacek-Larrondo\altaffilmark{15}}
\author{E. Leitch\altaffilmark{2,3}} %
\author{P. Mazzotta\altaffilmark{8}}
\author{C. Miller \altaffilmark{1}}
\author{S.J.C. Muchovej\altaffilmark{18}} 
\author{T. Schrabback\altaffilmark{19}}
\author{H.K.C. Yee\altaffilmark{20}}
\author{RCS-Team\altaffilmark{}}

\altaffiltext{1}{Department of Astronomy, University of Michigan, 1085 S. University Ave, Ann Arbor, MI 48109, USA} 
\altaffiltext{2}{Department of Astronomy and Astrophysics, The  University of Chicago, Chicago, IL 60637, USA}
\altaffiltext{3}{Kavli Institute for Cosmological Physics, The University of Chicago, Chicago, IL 60637, USA}
\altaffiltext{4}{Steward Observatory, University of Arizona, 933 North  Cherry Avenue, Tucson, AZ 85721, USA}
\altaffiltext{5}{Leiden Observatory, Leiden University, P.O. Box 9513, 2300 RA Leiden, The Netherlands}
\altaffiltext{6}{Department of Physics,450 Church St, University of Michigan, 500 Church Street, Ann Arbor, MI 48109, USA} 
\altaffiltext{7} {INAF-Osservatorio Astronomico of Trieste, via Tiepolo 11, 34121, Trieste, Italy}
\altaffiltext{8}{Dipartimento di Fisica, Universit\`a di Roma Tor Vergata, via della Ricerca Scientifica, I-00133, Roma, Italy} 
\altaffiltext{9}{Sustainable Engineering Group, 7475 Hubbard Avenue Suite 201, Middleton, WI 53562}
\altaffiltext{10}{Pontificia Universidad Cat\'{o}lica de Chile,  Santiago 22, Chile}
\altaffiltext{11} {Department of Physics, Harvard University, 17 Oxford Street, Cambridge, MA 02138} 
\altaffiltext{12} {Harvard-Smithsonian Center for Astrophysics, 60 Garden Street, Cambridge, MA 02138, USA} 
\altaffiltext{13}{South African Astronomical Observatory, P.O. Box 9, Observatory 7935, South Africa}
\altaffiltext{14}{Department of Physics and Astronomy, Johns Hopkins University, Baltimore, MD 21218, USA}
\altaffiltext{15} {Departement de Physique, Universite de Montreal,  C.P. 6128, Succ. Centre-Ville, Montreal, Quebec H3C 3J7, Canada} 
\altaffiltext{16} {National Research Council Fellow, National Academy of Sciences.} 
\altaffiltext{17} {U.S.\ Naval Research Laboratory, 4555 Overlook Ave SW, Washington, D.C.\ 20375, USA.} 
\altaffiltext{18} {California Institute of Technology - Owens Valley  Radio Observatory, Big Pine, CA 93513, USA} 
\altaffiltext{19} {Argelander Institute for Astronomy, University of Bonn, Auf dem H{\"u}gel 71, 53121 Bonn, Germany} 
\altaffiltext{20} {Department of Astronomy and Astrophysics,  University of Toronto, 50 St. George Street, Toronto, Ontario M5S  3H4, Canada}

\begin{abstract}
We present an initial study of the mass and evolutionary state of a massive and distant cluster,
RCS2~J232727.6-020437. 
This cluster, at z=0.6986, is the richest
cluster discovered in the RCS2 project. 
The mass measurements presented in this paper are derived from all
possible mass proxies:
X-ray measurements, 
weak-lensing shear,
strong lensing, 
Sunyaev Zel'dovich effect decrement, 
the velocity distribution of cluster member galaxies, 
and galaxy richness.
While each of these observables probe the mass of the cluster at 
a different radius, they all indicate that
\longuber~ is among the most massive clusters at this redshift, with
an estimated mass of $M_{200} \sim 3 \times10^{15}$\h\Msun.  
In this paper, we demonstrate that the various observables are all reasonably
consistent with each other to within their uncertainties.
\longuber~ appears to be well relaxed -- with
circular and concentric X-ray isophotes, with a cool core, and no indication of
significant substructure in extensive galaxy velocity data.
\end{abstract}


\section{Introduction}\label{s.intro}
High redshift clusters have been successfully identified in dedicated
surveys working with a range of cluster-selecting techniques and
wavelengths. These include cluster discoveries in deep
X-ray observations 
\citep[e.g.,][]{gio94, ros98, rom00, ebe01, rosati04, mullis05, sta06, rosati09}; 
optical$+$near 
infrared imaging \citep[e.g.,][]{gla05, stanford05,brodwin06, elston06,
  wit06, eisenhardt08, 
muzzin09, wil09, papovich10, brodwin11, santos11, gettings12,
stanford12, zeimann12, stanford14},
and detection of the the Sunyaev Zel'dovich (SZ) effect
\citep{staniszewski09,van10,williamson11,marriage11b,planck13,hasselfield13,reichardt13,brodwin14,bleem14}.

Nevertheless, despite this extensive effort, these surveys resulted in a modest number of high
redshift ($z\ga0.5$) and massive ($M\ga10^{15}$\Msun) galaxy clusters.
This relative
paucity of distant massive clusters is a reflection both of the
challenges inherent in detecting such clusters and of their intrinsic
rarity \citep[e.g.,][]{cro10}. Such clusters are the earliest and
largest collapsed halos; the observed density of distant massive
clusters is thus exquisitely sensitive to several cosmological
parameters \citep[e.g.,][]{eke96} and indeed the presence of a {\em
single} cluster in prior cluster surveys has been used to limit
cosmological models \citep[e.g.,][]{bah98}. Such clusters also offer, at least
in principle, the opportunity to test for non-gaussianity on cluster
scales if the cosmology is otherwise constrained
\citep[e.g.,][]{sar10}. 

At $0.6<z<1.0$, the most massive galaxy clusters known to date are 
CL~J1226$+$3332 \citep{maughan04} at $z=0.89$ with mass of
  $M_{200}=1.38\pm0.20\times10^{15}$\h\Msun~ \citep{jee09}, 
  ACT-CL~J0102--4915 at $z=0.87$ and with
  $M_{200}=2.16\pm0.32\times10^{15}$\h\Msun~
  \citep{menanteau12}, and MACS0744.8$+$3927 at $z=0.698$ with 
$M(<1.5\rm{Mpc})= 2.05\pm0.57\times10^{15}$\h\Msun~ \citep{applegate14}.
Recently discovered $z>1.0$ galaxy clusters
  appear to have more moderate masses,  e.g., 
  SPT-CL~J2106--5844 ($z=1.14$, $M_{200}=1.27\pm0.21\times10^{15}$\h\Msun;
  \citealt{foley11}),  SPT-CL~J2040--4451 ($z=1.48$,
  $M_{200}=5.8\pm1.4\times10^{14}$\h\Msun;
  \citealt{bayliss13}) and IDCS~J1426.5$+$3508 ($z=1.75$,
  $M_{200}=4.3\pm1.1\times10^{14}$\h\Msun;
  \citealt{brodwin12}).

Massive clusters at any redshift are amenable to detailed study with a
density of data that less massive systems do not present. The X-ray
luminosity of clusters scales as M$^{1.80}$ \citep{pra09}, the SZ
decrement as M$^{1.66}$ \citep{bon08}, the weak-lensing shear
approximately 
as $M_{200}^{2/3}$, and the galaxy richness in a fixed metric aperture (and hence
the available number of cluster galaxy targets for spectroscopic and
dynamical studies within a given field of view) scales as M$^{0.6}$
\citep{yee03} at these masses. Similarly it is expected that the most
massive clusters dominate the cross-section for cluster-scale strong
lensing \citep{hen07}. Thus the most massive clusters offer a wealth
of potentially well-measured observables which can be used, for
example, to study the correspondance between different mass proxies;
such study is critical to the success of surveys which aim to
use the redshift evolution of the cluster mass function as a
cosmological probe.

We present here detailed observations of a single massive cluster
selected from the Second Red-Sequence Cluster Survey
\citep[RCS2;][]{gilbank11}.  This cluster, \longuber~ (hereafter
RCS2327), was selected from \RCS~ in an early and partial cluster
catalog. Its optical properties indicated that it is a very massive
cluster, and justified an extensive followup campaign with
ground-based and space-based observatories at all
wavelengths, from X-ray to radio. 
Since its discovery, some of the properties of \clustername~ have been reported on in the
literature. \citet{gralla11} first measured its mass from
Sunyaev Zel'dovich array observations and its Einstein radius from
strong lens modeling. \clustername~ was rediscovered as the
highest significance cluster in the Atacama Cosmology Telescope survey
(Hasselfield et al. 2013), and  Menanteau et
al. (2012) also report on mass estimates from archival optical and
X-ray observations.

Although the discovery publication of \clustername~ has been
delayed, it was advertised in the past decade in various oral
presentations and conferences -- in
order to motivate more extensive followup effort by the
community. 
Indeed deeper and more detailed observations have been
conducted since, and will be the basis of future publications.  
This paper presents mass estimates
  from the initial survey and early multi-wavelength followup
  observations of \clustername, which collectively indicate that it is an
  unusually massive high-redshift cluster of galaxies.

This paper is organized as follows. 
The appearance of the cluster in the \RCS~ data and catalogs is discussed
in \S~\ref{s.rcs}. 
We
describe the various datasets and corresponding analyses (richness
and galaxy photometry, dynamics,
X-ray, SZ decrement, weak- and
strong-lensing) in detail in
\S~\ref{s.followup}. We discuss the implications of these observations
in \S~\ref{s.mass}  and conclude in
\S~\ref{s.conclusions}.

Throughout the paper we use the conventional notation $M_{200}$
($M_{500}$, $M_{2500}$) to denote the enclosed mass within a radius $R_{200}$
($R_{500},R_{2500}$), where the overdensity is 200 (500, 2500) the
critical matter density at the cluster redshift.
Unless otherwise stated, we used the WMAP 5-year cosmology
parameters \citep{kom08}, with 
$\Omega_{\Lambda}= 0.73$, $\Omega_{m} =0.27$, and $H_0 = 70$  km s$^{-1}$
\rm{Mpc}$^{-1}$. In this cosmology, $1\arcsec$ corresponds to 7.24~kpc at
the cluster redshift, $z=0.6986$.
Magnitudes are reported in the AB system.

\section{The Second Red-Sequence Cluster Survey and the Discovery of RCS2327}\label{s.rcs}

The Second Red-Sequence Cluster Survey (\RCS) is an imaging program
executed using the Megacam facility at CFHT.  RCS2 is described in full in
 \citet{gilbank11}. In short, images have been acquired
in the $g$, $r$, and $z$ filters, with integration times of 4, 8, and 6
minutes respectively, and all with sub-arcsecond seeing
conditions via observations in queue mode. The RCS2 data are
approximately 1-2 magnitudes deeper than the Sloan Digital Sky Survey 
imaging \citep{yor00}, with a 5-$\sigma$ point-source limiting magnitudes of 24.4, 24.3, 
and 22.8 mag in  $g$, $r$, and $z$ respectively. 
The
RCS2 survey data comprise 785 unique pointings of the nominally 1
square degree CFHT Megacam camera; the surveyed area is somewhat less
than 700 square degrees once data masking and pointing overlaps are
accounted for. The cluster and group catalog from RCS2 extend to
$z\sim1.1$, constructed using the techniques described in
\cite{gla05}. The $g$-band imaging improves the overall performance
at lower redshifts (compared to RCS1; Gladders \& Yee 2005), and
makes the survey more adept at detecting strong 
lensing clusters, since lensed sources tend to have blue colors. 

\begin{figure*}
\centering
\includegraphics[scale=1]{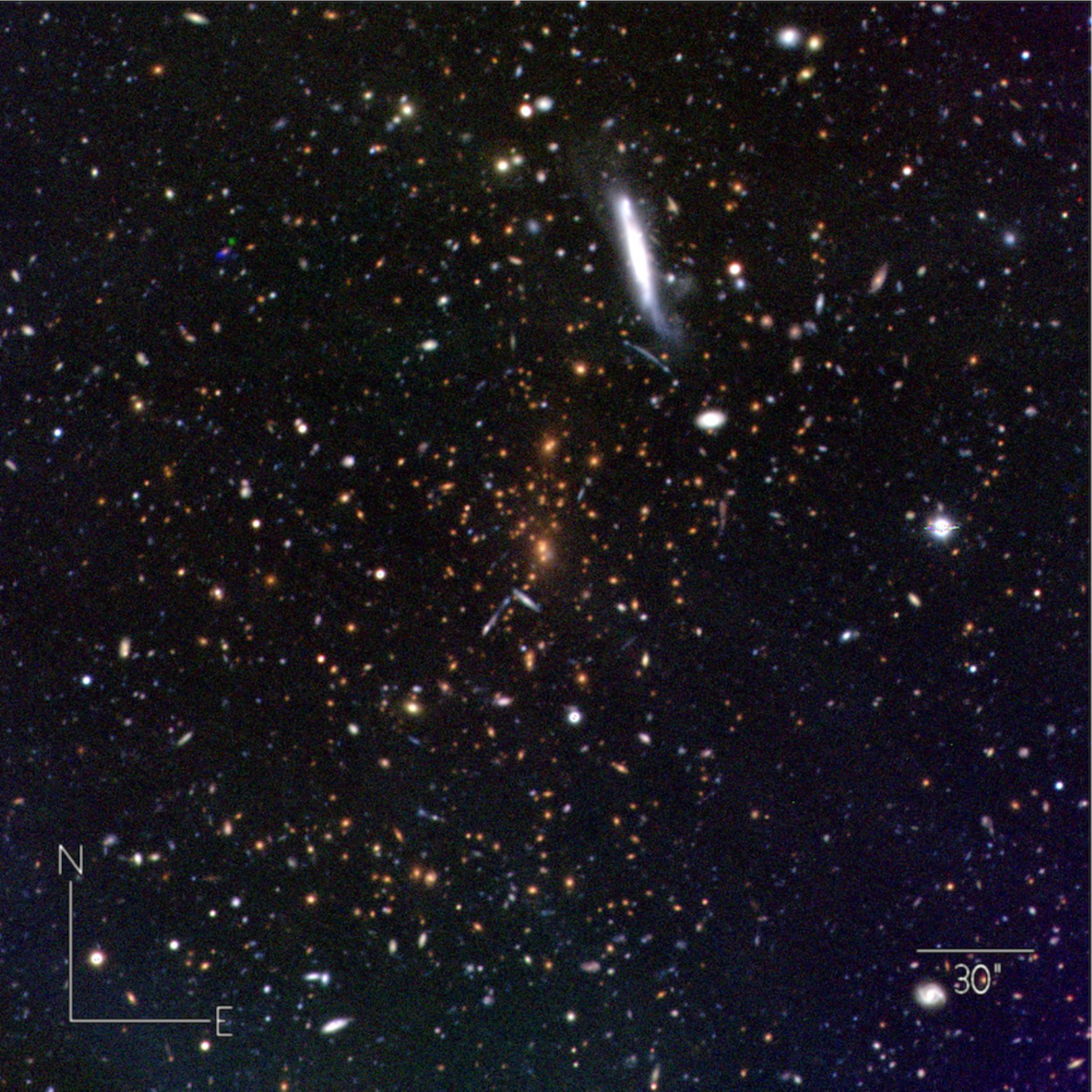} 
\caption{A composite $gri$
color image of \uber~ from imaging from the LDSS-3 instrument on
Magellan (see \S~\ref{s.richness}). The field of view is 2x2 proper Mpc at the
redshift of the cluster, centered on the BCG. RCS2327 is obviously
demarcated by the abundance of red early-type galaxies in the center
of the image. Immediately south of the foreground bright galaxy to the
north-northeast is an obvious strongly lensed merging double image
(see Section~\ref{s.sl}) which was also readily apparent in the original RCS2
survey data.}
\label{colim}
\end{figure*}

\begin{figure}
\centering
\includegraphics[scale=0.5]{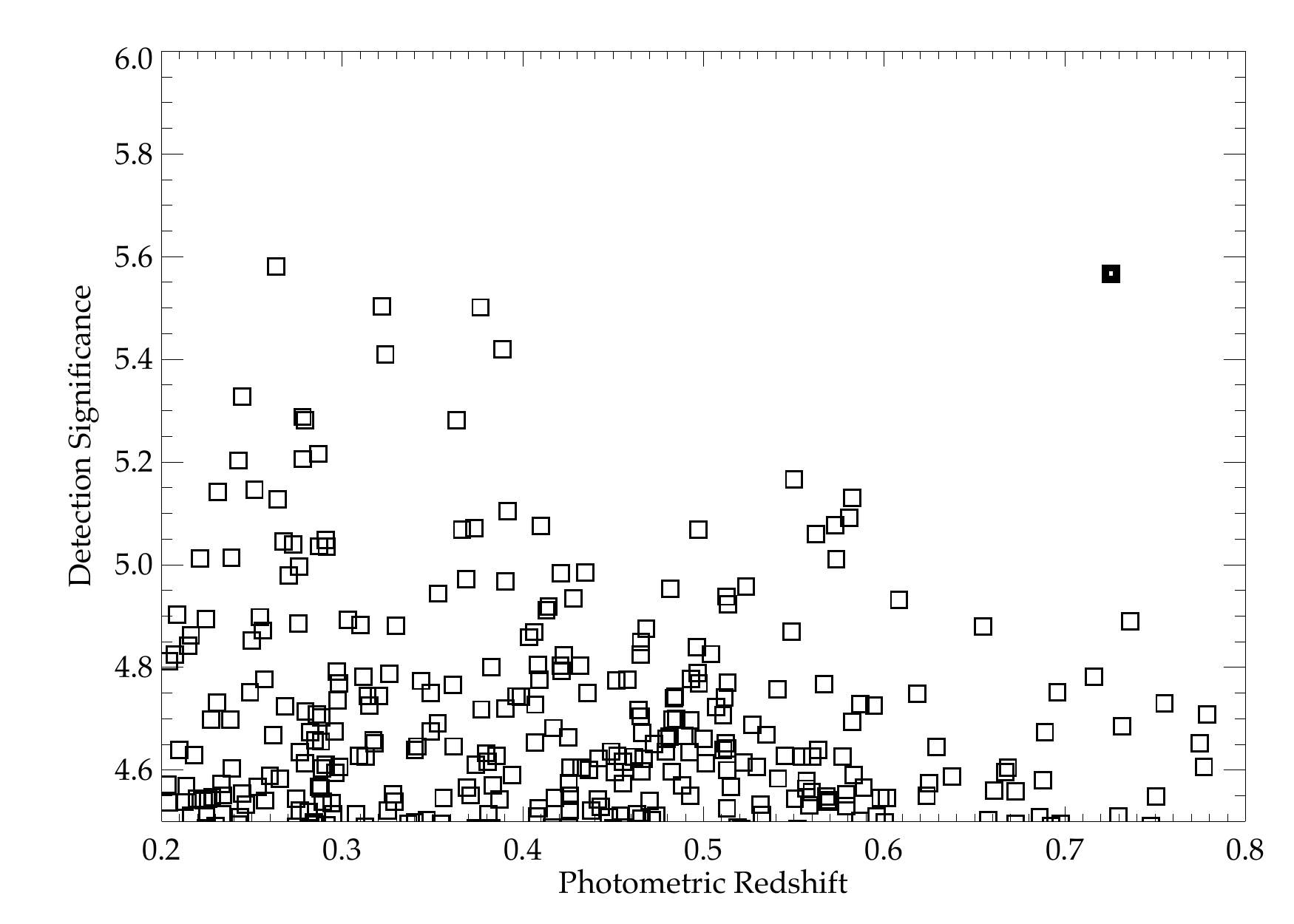} 
\caption{The detection significance
versus redshift of RCS2327 compared to the distribution of these
properties for clusters from the RCS2 cluster catalog. RCS2327 is
indicated as a heavy symbol and is an obvious outlier; it is as
significant as any cluster detected at lower redshifts.}
\label{ubercon}
\end{figure}

RCS2327 was discovered in 2005 in an early and partial version of the RCS2
cluster catalog. An examination of the RCS2 survey images made it
clear that it was an unusually massive object. A color image of
RCS2327 is shown in Figure \ref{colim}. The original RCS2 imaging
data clearly showed at least one strongly lensed arc, and the
indicated cluster photometric redshift was $z\sim0.7$. A plot of the detection
significance versus photometric redshift for clusters from the RCS2
cluster catalog is shown in Figure \ref{ubercon}. The RCS2 imaging
data are fairly uniform, and so at a given redshift the detection
significance is a meaningful quantity that is not strongly affected by
data quality from region to region of the survey. At high redshifts
RCS2327 is the most significant cluster detected. Furthermore, a
cluster of a given richness and compactness (both of which influence
detection significance) will be detected as a more significant
object at lower redshifts; the fact that RCS2327 is detected with a
significance as great as any lower redshift clusters implies that it
is likely the most massive cluster in this sample. Even from these
basic data and considering the volume probed it is apparent that
RCS2327 is a remarkably massive cluster, worthy of significant
followup.

The cluster is located at R.A.$=$23:27:27.61 (J2000) and
Decl.$=-$02:04:37.2 (J2000); this is the position of the brightest
cluster galaxy (BCG) and is coincident with the center of the cluster
X-ray emission (see \S~\ref{s.xray} below).

\section{Followup Observations and Mass Estimates}\label{s.followup}
In this section we describe the multi-wavelength followup
observations of \clustername. Based on these observations, we are able
to estimate the mass of \clustername~ from 
richness, galaxy dynamics, X-ray, Sunyaev Zel'dovich effect, weak lensing and
strong lensing. The different mass proxies naturally measure either
spherical mass or a projected mass along the line of sight
(usually referred to as cylindrical mass or aperture mass).  
Moreover, each mass proxy is sensitive to mass at a different radial scale: 
strong lensing measures the projected mass density
at the innermost parts of the cluster, typically $\sim$100-500 kpc, and is insensitive to the mass
distribution in the outskirts; SZ decrement and X-ray measure the
mass at larger radii (typically R$_{2500}$) and lack the resolution at
the center of the cluster; weak lensing reconstructs the projected 
mass density out to R$_{200}$, with poor resolution at the center as
well. 
Dynamical mass (from the velocity distribution of cluster galaxies) is
used to estimate the virial mass. 
We note that these mass proxies are
not always independent, and rely on scaling relations and
assumptions.      
In the following subsections, we describe the data and our analysis to derive the
cluster mass from each mass proxy.
In \S~\ref{s.mass} we
compare the masses derived from the different mass proxies. 

\subsection{Deep Multi-color Imaging, Galaxy Distribution and Richness}\label{s.richness}

In addition to the RCS2 survey imaging data, available imaging data
on RCS2327 includes images from the LDSS-3 imaging spectrograph on the
6.5m Clay telescope, taken during a run in September 2005. Total
integration times were 16, 12, and 10 minutes in the $g$,$r$ and $i$
filters respectively. The point-spread-function width at half maximum
in the final stacked images is 0\farcs60 ($i$), 0\farcs65 ($r$), and 0\farcs80
($g$) with some image elongation due to wind shake present principally
in the bluest band. These data cover a circular field of view 8' 
in diameter, centered on the BCG. Figure \ref{colim} is
constructed from these data.

The RCS2 data are best suited to measurements of cluster richness, as
they are well calibrated and naturally include excellent background
data, and are readily connected to the cosmological context and
calibration of the mass-richness relation provided by the RCS1
program \citep{gla05,gla07}. The multi-band LDSS-3 images are deeper,
with better seeing than the RCS2 images, and we use these data for
computing detailed photometric properties - principally
color-magnitude diagrams. For this we focus our analysis below on the
$r$ and $i$ observations, since this filter pair has the best image
quality, and almost perfectly straddles the 4000\AA~break at the cluster
redshift.

\begin{figure}[h]
\centering
\includegraphics[scale=1]{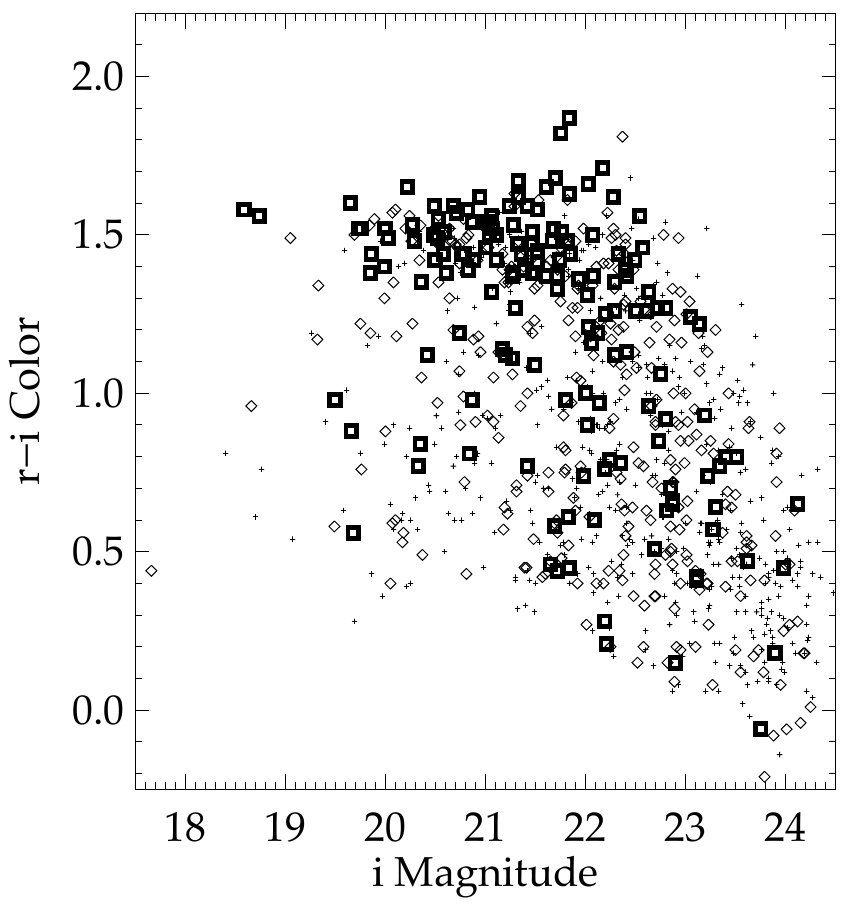} 
\caption{The color-magnitude diagram for a 1
Mpc radius field of view centered on RCS2327, in the $r$ and $i$
filters. Galaxies have been divided into three radials bins of equal
radius; galaxies in the outermost bin are plotted as small pluses, and
those in the central bin as heavy squares. Only galaxies with
photometric uncertainties in both filters of less than 0.2 magnitudes are shown.}
\label{cmrphot}
\end{figure}

\begin{figure}
\centering
\includegraphics[scale=1]{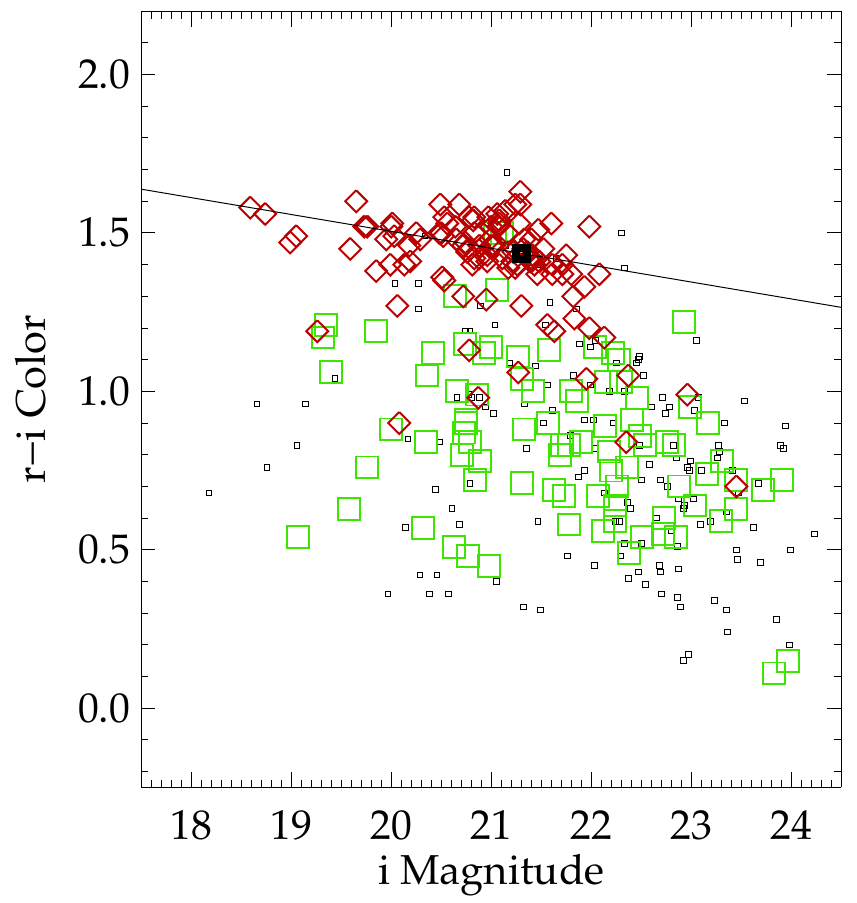} 
\caption{The color-magnitude diagram for
galaxies with spectroscopic redshifts, regardless of position, in the
$r$ and $i$ filters. Spectroscopic early-type cluster members are
shown as red diamonds; other cluster members are shown as green
squares. Non-members are shown as small squares. Only galaxies with
uncertainties in both filters of less than 0.2 magnitudes are
shown. The best fitting red-sequence model, derived as described in
\S~\ref{s.richness}, is overplotted with the characteristic magnitude $i^*$ indicated
by the heavy black square.}
\label{cmrspecphot}
\end{figure}

Figure \ref{cmrphot} shows an $r-i$ color-magnitude diagram of all
galaxies at projected radii less than 1 Mpc from the cluster
center. The red sequence of early-type cluster members is obvious,
emphasizing the extraordinary richness of this cluster in comparison
to most other clusters in the literature at a similar redshift
\citep[e.g.,][]{del07,gla05}.  Figure \ref{cmrspecphot} shows only
galaxies for which a spectroscopic redshift is available, plotted by
SED type. As expected, galaxies which are both cluster members and
have early-type spectra are almost all red-sequence members. From the
spectroscopically confirmed early-type cluster galaxies, with simple
iterated 3-$\sigma$ clipping \citep[e.g., as in][]{gla98}, we fit a
linear red sequence relation, given by
$(r-i)=-0.053\times(i-i^*)+1.436$. We take the characteristic
magnitude for cluster galaxies in RCS2327 as $i^*=21.3$, consistent to
within 0.05 magnitudes with the models in both \cite{gla05} and
\cite{koe07}.  These models include a correction for passive
evolution.  The measured scatter of early-type galaxies about the best
fit red sequence is less than 0.05, consistent with that seen in other
rich clusters at a range of redshifts \citep{hao09,mei09}. The
best fitting model is indicated in Figure \ref{cmrspecphot}. These data
demonstrate that RCS2327 appears as expected for a well-formed
high-redshift cluster, albeit an extraordinarily rich example.

We derive a total richness for RCS2327 of $B_{gcT}$=3271$\pm$488
Mpc$_{50}^{1.8}$, and a corresponding red-sequence richness of
$B_{gcRS}$=2590$\pm$413 Mpc$_{50}^{1.8}$ \citep[see][for a detailed
explanation of the $B_{gc}$ parameter]{gla05}. Calibrations relevant
to the measurement of this richness have been taken from the RCS1
survey, which also uses the $z'$-band as the reddest survey filter. A
direct comparison of the total richness to the scaling relations in
\cite{yee03} nominally corresponds to a mass of
M$_{200}=4.6_{-1.1}^{+1.2} \times 10^{15}$ \h\Msun~ with a
significant uncertainty, given the known scatter in richness
as a mass proxy \citep{gla07,roz09}, and the lack of direct
calibration of the richness-mass relation at the redshift of RCS2327.
Furthermore, the relevant richness to use in comparison to the scaling
relation in \cite{yee03} is not obvious; though the richness values in
\cite{yee03} are for all galaxies, the small blue fraction in that
sample and the significant observed evolution in the general cluster
blue fraction \citep{loh08} from the redshift of that sample (mean
z=0.32) to the redshift of RCS2327 suggests that the (less evolving)
red-sequence richness may be a more appropriate measure. With that in
mind we note that the mass corresponding to $B_{gcRS}$ is
M$_{200}=3.2_{-0.8}^{+0.9} \times 10^{15}$\h\Msun.
Given the limitations of this analysis however, we do not use a
richness-derived mass extensively in the analysis in \S~\ref{s.mass}, but simply
note here that RCS2327 is remarkably rich.

\begin{figure*}
\centering
\includegraphics[scale=0.75]{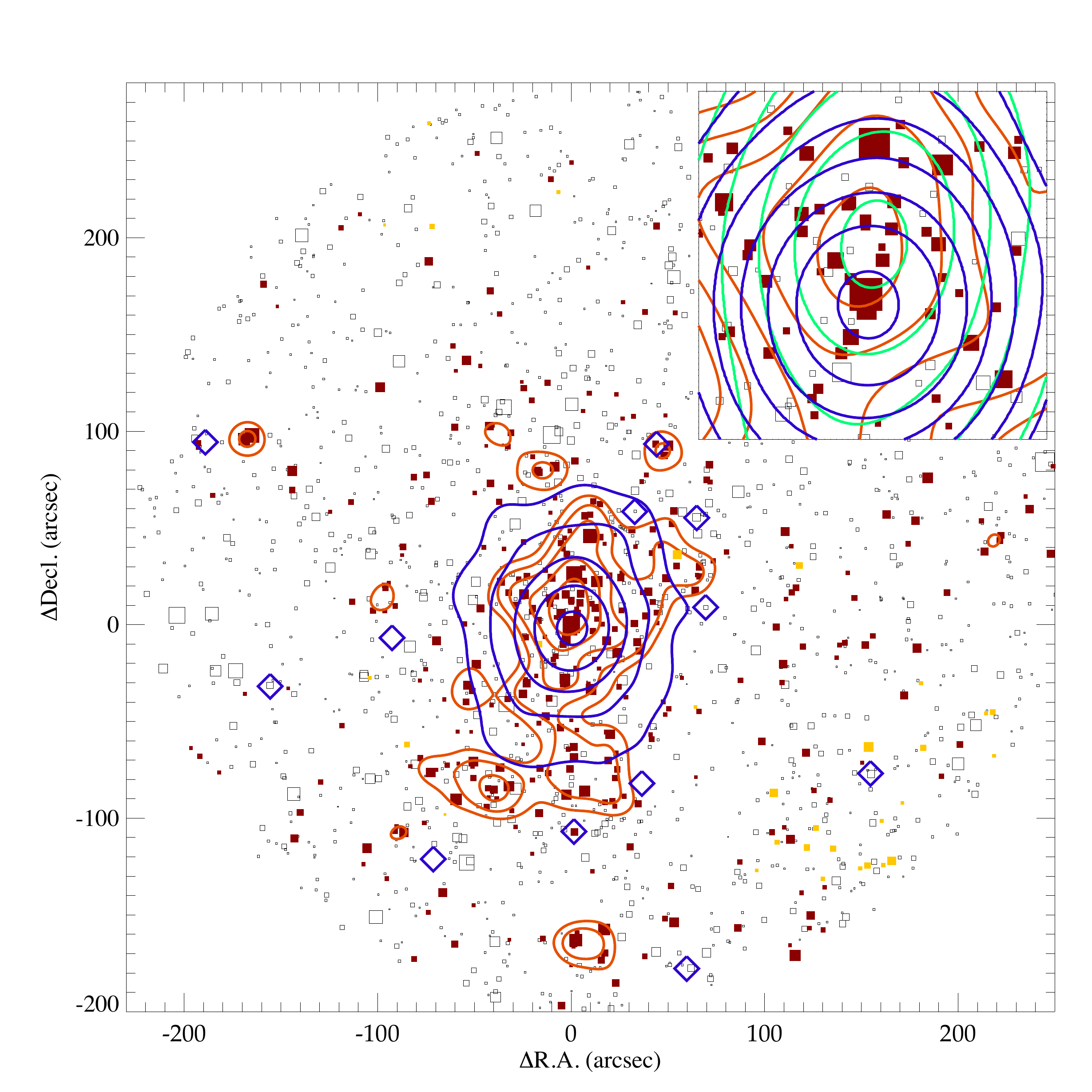} 
\caption{ The distribution of
red-sequence galaxy light (red contours) smoothed with a 150 kpc FWHM
gaussian, and the full band X-ray light (blue contours) similarly
smoothed.  
Blue diamonds mark X-ray point sources; the correspondance
between some of these sources to optical counterparts validates the
astrometric matching between these data. Individual galaxies are
indicated by squares, with symbol size proportional to brightness;
solid red squares have colors consistent with the cluster red-sequence
and are the basis for the overplotted contours, and solid yellow
squares show spectroscopically confirmed members of the secondary
structure noted in Figures \ref{dynhist} and \ref{dynrad} and
\S~\ref{s.spec}. The inset shows the central 1\arcmin$\times$1\arcmin region
with the same data, with the strong lensing main halo mass distribution
 also indicated (green contours). The
contour levels have been chosen to highlight the core position and
outer shape of the light distributions.}
\label{overlay}
\end{figure*}

\subsection{Optical Spectroscopy} \label{s.spec}
Spectroscopic observation of galaxies in the field of RCS2327 has been conducted
using the Magellan telescopes. Data were acquired in both normal and
nod-and-shuffle modes using LDSS-3, during runs in August and
November, 2006, and a total of 3 masks
with the GISMO instrument in June 2008. RCS2327 was
also observed using the GMOS instrument on the Gemini South telescope
in queue mode in semester 2007B, yielding redshifts of potential
lensed sources; the Gemini data are discussed in more detail in 
\S~\ref{s.sl} below.

All Magellan spectra have been reduced using standard techniques, as
implemented in the COSMOS
pipeline\footnote{http://obs.carnegiescience.edu/Code/cosmos}. The
bulk of the LDSS-3 observations (apart from a single early mask, which
established the cluster redshift at $z\sim0.70$) were acquired using a
6000\AA-7000\AA~band-limiting filter; this allows for a high density
of slits, at the expense of a significant redshift failure rate
(specifically, \OII~  is undetectable outside of
$0.61<z<0.87$, and the Ca H and K lines are undetectable outside of
$0.53<z<0.76$). The GISMO observations were conducted using a band
limiting filter covering 5700\AA-9800\AA. 

A total of 353 robust redshifts were measured from these
data. Most are unique, with overlap between observations with
different instruments or runs amounting to a few galaxies per
mask. From six galaxies in common beween the LDSS-3 and GISMO data the
mean difference in redshifts is measured to be 135 km~sec$^{-1}$, and
the uncertainty within observations using a single instrument is
measured to be less than 100 km~sec$^{-1}$. Neither of these
uncertainties is significant in the analysis below. Redshifts were
measured using a combination of cross correlation and line measurement
techniques, and cross correlation measurements of absorption systems
were only retained if (at minimum) the H and K lines were individually
visible. Apart from possible mis-interpreted single emission line
redshifts in the LDSS-3 spectra, the measured redshifts are
robust. Each spectrum was also classified as either an emission or
absorption type, with post-starburst (showing strong Balmer lines) or
AGN features also noted when present.

\begin{figure}
\centering
\includegraphics[scale=0.88]{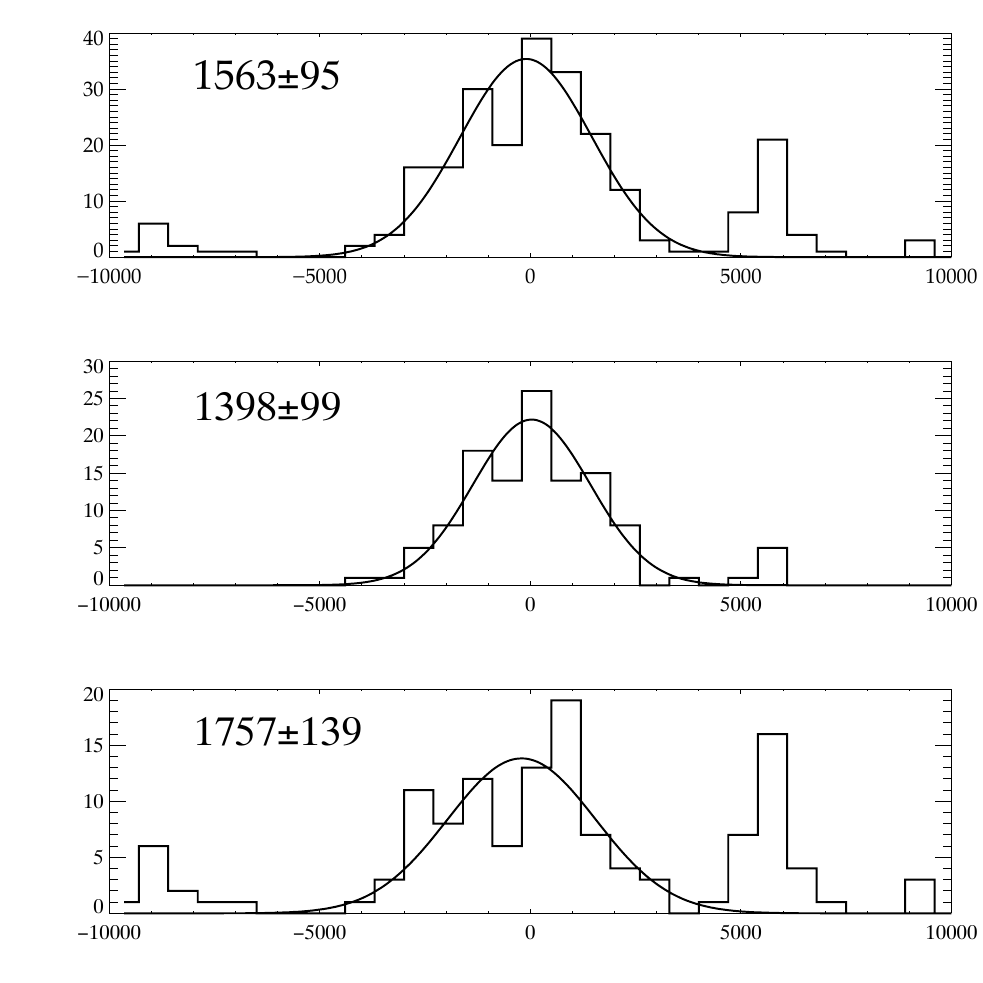}
\caption{The velocity distribution of galaxies about the mean cluster
redshift for all galaxies (top), absorption line only galaxies
(middle), and all emission line galaxies (bottom). Velocity
dispersions in the restframe in km~sec$^{-1}$are as indicated.}
\label{dynhist}
\end{figure}

A histogram of the galaxy velocities around the mean cluster redshift
of 0.6986$\pm$0.0005 is shown in Figure \ref{dynhist}. The measured
velocity dispersion for 195 cluster members is 1563$\pm$95
km~sec$^{-1}$ with uncertainties measured from a bootstrap
analysis. The cluster is well separated from other structures. The
velocity dispersion using only the 110 galaxies with early type
spectra is 1398$\pm$99 km~sec$^{-1}$, and similarly using all other
cluster members we derive 1757$\pm$139 km~sec$^{-1}$ -- a factor of
1.27$\pm$0.14 larger. These differences are as expected and in line
with that observed for relaxed X-ray selected clusters at lower
redshifts, where the typical ratio in velocity dispersion of blue to
red cluster members is 1.31$\pm$0.13 \citep{car97}.

\begin{figure}
\centering
\includegraphics[scale=0.51]{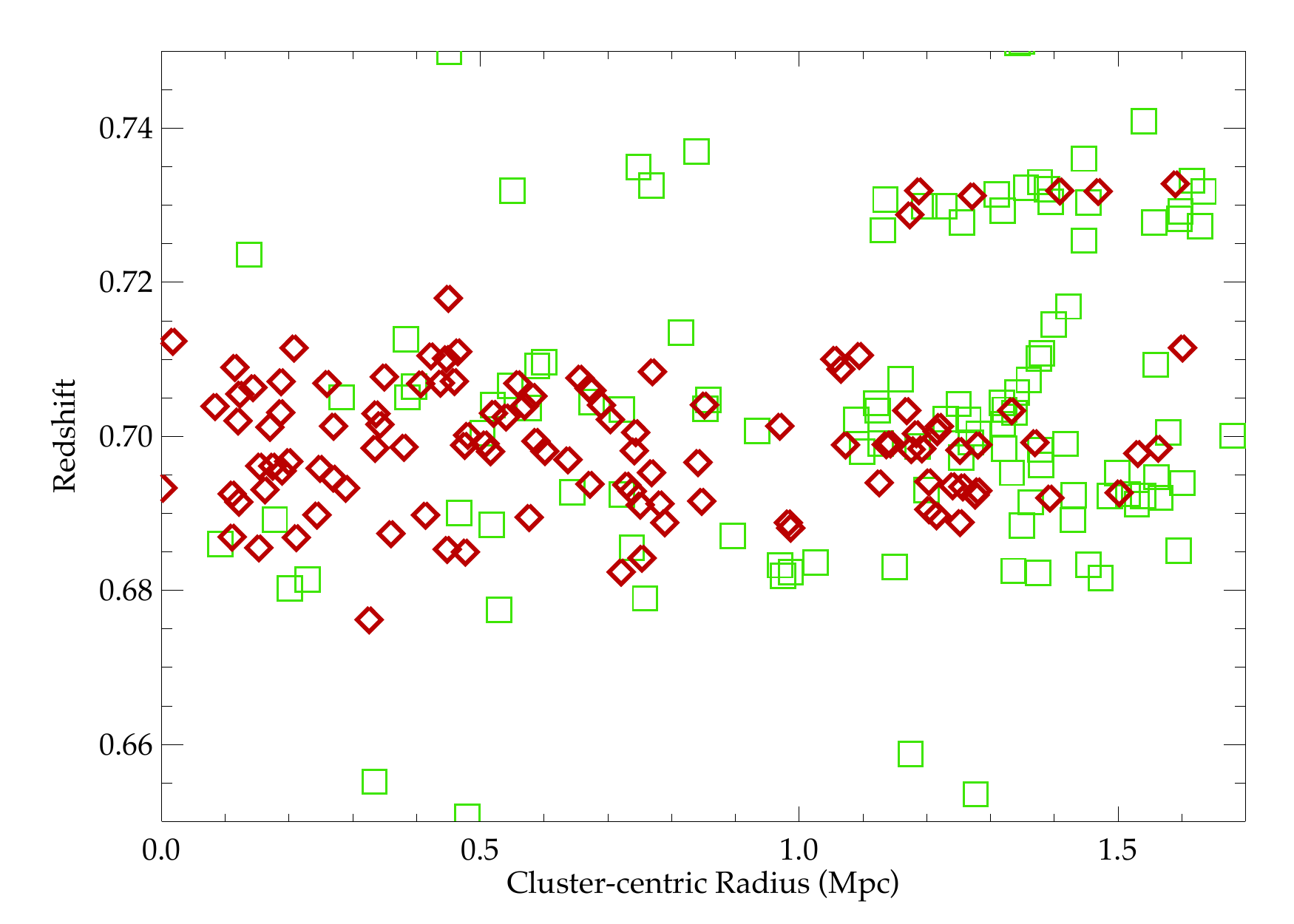}
\caption{Redshifts versus cluster-centric radius for
absorption-line-only galaxies (red diamonds) and all other galaxies
(green squares) for all galaxies within $\Delta z=0.05$ of the
RCS2327. The field of view of the GISMO instrument is approximately 0.7
Mpc in these coordinates; the denser sampling this provides in the
cluster core, combined with the details of slit placements in the
LDSS-3 instrument yields a complex slit weighting with radius which is
responsible for the apparent deficit of galaxies at $\sim$1 Mpc.}
\label{dynrad}
\end{figure}

The velocity distribution of cluster members versus projected radius,
in Figure \ref{dynrad}, shows several trends also consistent with that
expected for a relaxed cluster. The velocity dispersion is a declining
function of cluster-centric radius, an effect most apparent in the
early-type galaxies. In non-overlapping radial bins of 0.5 Mpc in
radius, and at mean radii of 0.27, 0.71 and 1.27 Mpc, 
we find velocity
dispersions of 1626$\pm$127 km~sec$^{-1}$, 1268$\pm$147 km~sec$^{-1}$,
and 1034$\pm$201 km~sec$^{-1}$, respectively. The radial distribution of emission
line members relative to absorption line cluster members is also as
expected, with proportionately more actively star-forming systems
found at large radii. A clear interpretation of this result is difficult
given the complexity of the sampling from multiple masks from multiple
instruments with differing fields of view, and the weighting of slit
assignments toward photometric red-sequence members; the data shown in
Figure \ref{dynrad} are at least consistent with
expectations. Finally, a KS test of the velocity distribution shows at
best marginal evidence for velocity substructure, with the velocity
distribution inconsistent with a normal distribution at a modest 1.3
sigma using all galaxies. Using only early type members, there is not
even marginal evidence for velocity substructure.

We also note the presence of a secondary structure separated from the
main cluster by 5700 km~sec$^{-1}$. This structure is dominated by
emission line galaxies, has a velocity dispersion of 400
km~sec$^{-1}$, and is located to the edge of the spectroscopic field
of view, as can been seen in Figures \ref{overlay}, \ref{dynhist}, and
\ref{dynrad}. It is not significant for any of the analyses below.

\subsubsection{Dynamical Mass Estimates from Velocity
  Dispersion}\label{s.veldisp}
The observed velocity dispersion is converted to mass through the 
virial scaling relation derived from simulations \citep{evr08}. This 
relationship is described as
\begin{equation}
M_{200} = \frac{10^{15}}{h(z)}\left(\frac{\sigma_{obs}}{\sigma_{15}}\right)^\alpha,
\end{equation}
where $\sigma_{obs}$ is the observed 1-d velocity dispersion of the 
cluster, $\sigma_{15}$ is the velocity dispersion normalized for a 
$10^{15}$\msun~ cluster, and $\alpha$ is the slope of the scaling 
relation. \citet{evr08} find the best fit parameters for a multitude 
of cosmologies and velocity dispersions measured from dark matter 
particles to be $\sigma_{15} = 1082.9$ km s$^{-1}$ and $\alpha = 2.975$. 
Using this scaling relation with the measured velocity dispersion for 
all galaxies, the resulting mass is $M_{200} = 2.97^{+1.40}_{-0.95} 
\times 10^{15}$\h\msun. The uncertainties in mass assume 
a $13\%$ total uncertainty in velocity dispersion. This included 
both the statistical uncertainty, which is small for the large number 
of galaxies observed in this sample, and the systematic 
uncertainty, which includes line-of-sight effects, cluster 
shape/triaxiality, and foreground/background contamination 
(Gifford et al. 2013, Saro et al. 2013). Saro et al. (2013) re-fit the 
scaling relation to a semi-analytic galaxy catalog for the Millennium 
Simulation and find the parameters to be $\sigma_{15} = 938.0$ km s$^{-1}$ 
and $\alpha = 2.91$. The resulting mass using these parameters is 
$M_{200} = 3.08^{+1.42}_{-0.97} \times 10^{15}$\h\msun.

\subsubsection{Dynamical Mass Estimates from the Caustic Method}\label{s.caustics}
The distribution of radial velocities of cluster galaxies as a
function of cluster-centric radius can be used to estimate its mass
using the caustic technique (Diaferio \& Geller 1997, Gifford et
al. 2013, Gifford \& Miller 2013). 
This method relies on the expectation
that cluster galaxies that have not
escaped the potential well of the cluster halo occupy a
well-defined region in a radius-velocity phase space confined by the
escape velocity from that potential, $v_{esc}(r)$.
We follow the techniques outlined in 
~\citet{Gifford13}, and refer to that publication for a full
description of the methods applied here. 

\begin{figure}
\centering
\includegraphics[scale=0.45]{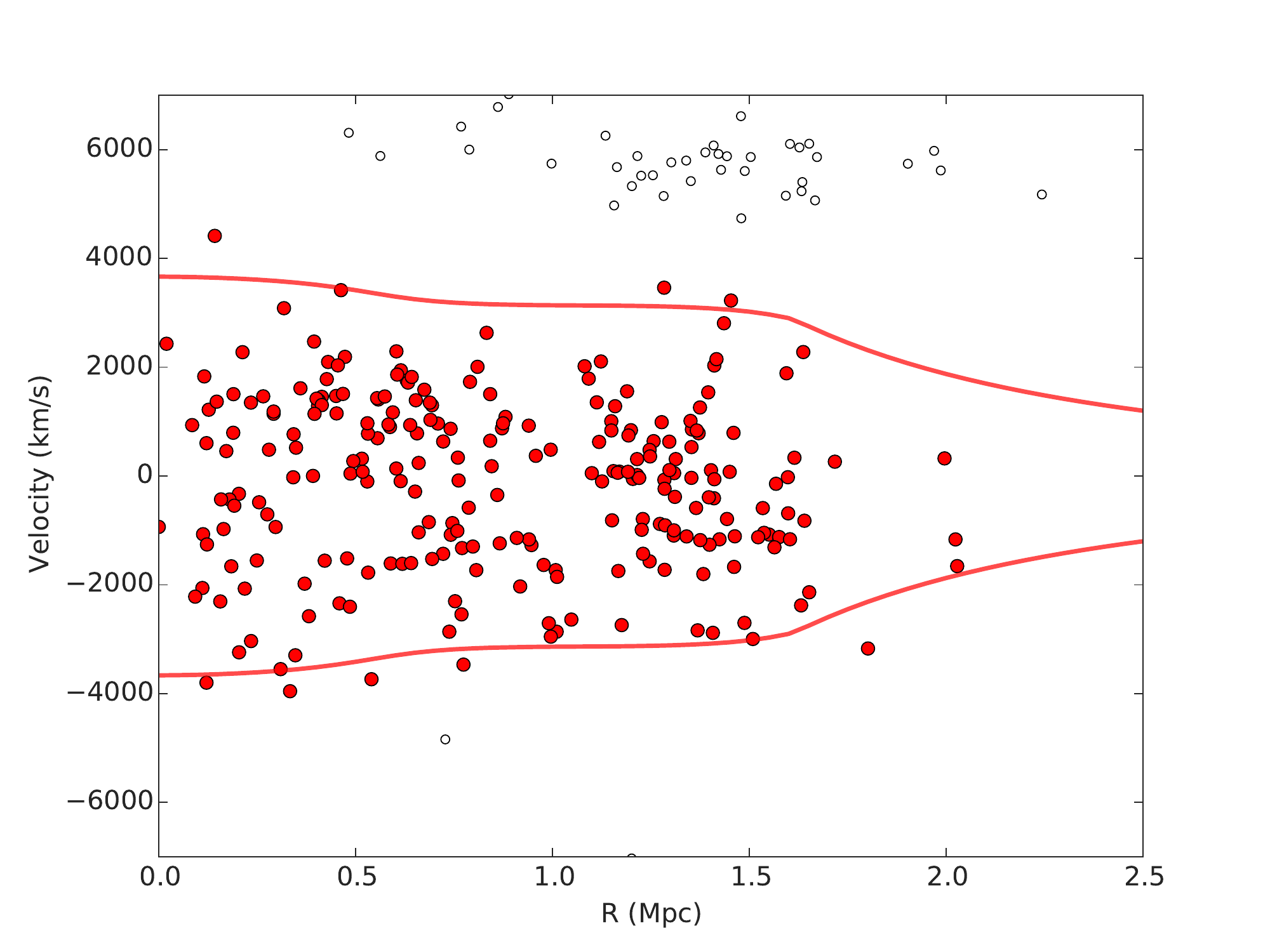} 
\caption{ The projected radius vs line-of-sight velocity phase space
  for galaxies identified as members by the caustic algorithm (red
  circles) and non-members (open circles). The estimated escape
  velocity surfaces (or caustics) are represented by solid red lines
  which are symmetric with respect to the cluster velocity. }
\label{fig.caustics}
\end{figure}

Figure~\ref{fig.caustics} shows the radius-velocity space of $\sim 250$
galaxies. We fit an iso-density contour to the data to find $v_{esc}$
as indicated by the velocity edge in phase-space density. 
The enclosed mass can be derived as
\begin{equation}
G M_{200} = \int^{r_{200}}_0 \mathcal{F}_{\beta} (r) \mathcal{A}^2(r) dr,
\end{equation}
where $ \mathcal{A}^2(r)$ is the square of the line-of-sight escape
velocity, and $ \mathcal{F}_{\beta} (r)$ is a function of the 
potential, density, and velocity anisotropy, corrected for projection effects. 
We apply the common convention of assuming that $\mathcal{F}_\beta$ is
 constant. Physically, this parameter depends on the unknown
concentration and velocity anisotropy profile of the
cluster. Disagreement on the constant value of $\mathcal{F}_\beta$
that results in unbiased mass estimates on average persists in the
literature with values ranging from 0.5 to 0.7. \citet{Gifford13}
find that $\mathcal{F}_\beta = 0.65$ results in mass estimates with less
than 4\% mass bias for several semi-analytic catalogs available for
the Millennium Simulation, and we adopt this value for this study.

We derive a dynamical mass of $M_{200}= 2.94^{+1.03}_{-0.76} 
\times 10^{15}$\h\msun. The uncertainty in
the derived mass using the caustics technique depends on the number of
galaxies used; from caustic mass analysis of the Millenium Simulation semi-analytic galaxy
catalogs, \citet{Gifford13} find that for $N_{gal} \geq 150$ the scatter
is $\leq 30\%$ with a bias of $< 4\%$.

{ We note that the velocity dispersion of galaxies identified as possible members by
the caustic technique is $1586\pm58$ km s$^{-1}$, in agreement with the estimates
in \S~\ref{s.spec}}. 

\subsection{{\it Chandra} Observations, X-ray properties and Mass Estimates} \label{s.xray}

RCS2327 was observed on two separate occasions with the {\it Chandra
  X-ray Observatory}. A first 25 ks observation was carried out on
2007 August 12 (Cycle 8 Proposal 08801039; PI: Gladders) using the
ACIS-S array. The early analysis suggested that the
cluster was massive, X-ray regular, and possibly hosting a cool
core. 
These hints justified the need of deeper data obtained in 2011
with ACIS-I (Cycle 13 Proposal 13800830; PI: Hicks). 

The two deep Cycle 13 ACIS-I pointings (150 ks total) sample
a more extended field around the cluster,
and result in a higher signal-to-noise ratio than the Cycle 8 observation. 
Since the background is better
understood than that of the ACIS-S configuration, combined with the
very small increase in signal-to-noise ratio that would be gained by
combining both datasets, we chose to analyse the 2011 data separately
and not co-add the two epochs. 

\begin{figure}
\centering
\includegraphics[scale=0.5]{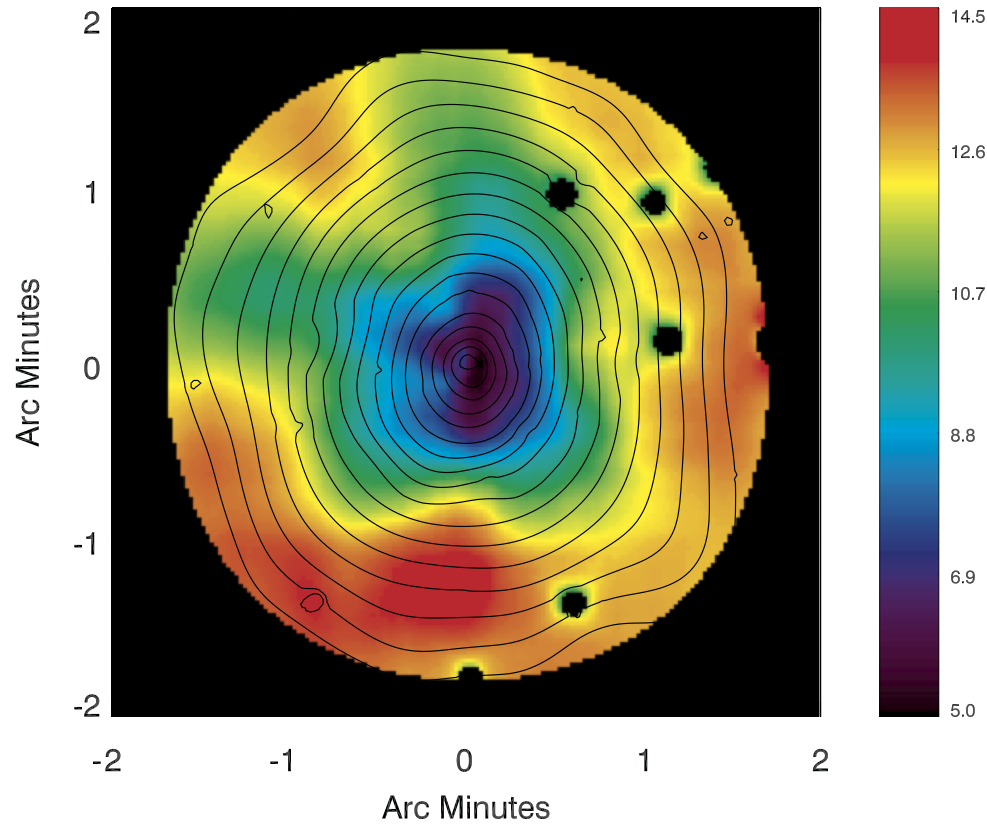} 
\caption{ A 4'$\times$4' temperature map of \uber, derived from the
  deep  {\it Chandra} Cycle 13 data. The image is centered on
  RA=23:27:27.53, Dec=$-02$:04:35.6. The temperature is indicated by the
  color scale, in keV. The contours are X-ray flux isophotes. Point sources have been
  masked. The temperature and flux maps clearly indicate that \uber~ has a cool
  core, and a regular morphology, with no significant substructure.}
\label{ktmap}
\end{figure}

The X-ray data reduction follows Martino et al. (2014) and Bartalucci et
al. (2014), with minor modifications. 
We use both the count statistics and
the control on systematics offered by a multi-component modeling of
the background noise \citep[see][for
details]{bartalucci.etal.14}. Filtering the hard and soft event
light curve reduces the total exposure time by 10\% to $\sim130$
ks.  We bin the photon events in sky coordinates with a fixed angular
resolution of 1\farcs4 and a variable energy
resolution that matches the detector response. The effective exposure
time and the estimated background noise level are similarly
binned. Following the {\it Chandra} Calibration database CALDB
4.6.1, when computing the effective exposure we take into account the spatially
variable mirror effective area, quantum efficiency of the detector,
CCD gaps, bad pixels, and a correction for the motion of the telescope.
Our background noise model includes Galactic foreground, 
cosmic X-ray background, and false detections due to cosmic ray
induced particles. For the particle background spectrum, we 
use the analytical model proposed by \cite{bartalucci.etal.14}.  
The amplitude of all the other components was determined from the data
outside the region of the field of view covered by the target. 
We derive the temperature map following techniques described in \citet{bourdin_mazzotta08}.
Figure~\ref{ktmap} shows the X-ray flux isophotes overplotted on the
false-color temperature map. The cluster has a regular X-ray
morphology and does not show significant substructure or X-ray
cavities \citep{hlavacek14}. 

\subsubsection{X-ray surface brightness, temperature, and metallicity profiles}
We extract the surface brightness profile (Figure~\ref{xrayprof}a) from an effective exposure
and background-corrected soft band ([0.5-2.5]
keV) image, after excluding point sources. The profile averages the
surface brightness in concentric annulli centered on the maximum of a
wavelet-filtered image of the cluster. 
The temperature and metallicity profiles (Figure~\ref{xrayprof}b,c) were
calculated in five radial bins out to $\sim 840$
kpc, each containing at least 2000 counts in the [0.7-5] keV
band. 
The measurements of temperature and metallicity assume
redshifted and absorbed emission spectra modeled with the
Astrophysical Plasma Emission Code (APEC, Smith et al. 2001), adopting the
element abundances of \cite{gre.sau98} and neutral hydrogen
absorption cross sections of \cite{balu.mcc92}. The spectra, modified
by the effective exposure and background, are also convolved with a
function of the redistribution of the photon energies by the detector. The
assumed column density value is fixed at $4.73 \times 10^{20}$
cm$^{2}$ from measurements obtained near our target by the 
Leiden/Argentine/Bonn (LAB) Survey of galactic HI \citep{kalberla.etal.05}. The
redshift is fixed to $z=0.6986$.

We find that the metallicity increases from $\sim0.2$
solar at \rtfh~ to $\sim0.6$ solar at the cluster core.
\uber~ shows a temperature gradient towards the center of the cluster,
indicating a significant cool core.  The temperature in the estimated $[0.15-1] \times R_{500}$ region (roughly
out to 1Mpc, see \S~\ref{s.xraymass}) is
$T_X= 13.9_{-1.8}^{+2.4}$ keV. 
We also estimate the cooling times as a function of cluster-centric
radius (following the prescription described in \citealt{hlavacek13}), and find that the cooling time
profile decreases mildly towards the center from $\sim40$ Gyr at 400 kpc, to
$\sim4$ Gyr at the core.  From the gas density (\S~\ref{s.xraymass})
and temperature profile, we find that the central entropy is $46 \pm28$ keV cm$^{-2}$.

\subsubsection{X-ray Mass Estimate.}\label{s.xraymass}
To measure the X-ray mass we follow the {\it forward} procedure
described in \cite{meneghetti10} and \cite{rasia.etal.12}. In
short, analytic models are fitted to the projected surface density and
temperature profiles and subsequently analytically de-projected. The
3D information are then folded into the hydrostatic mass equation
\citep{vikhlinin06}. 
The surface brightness is parametrized via a modified $\beta-$model
with a power-law trend in the center and a steepening
behavior in the outskirts, plus a second $\beta-$model to describe the
core: 
\begin{equation}
n_p n_e (r) = n_0^2\frac{(r/r_c)^{-\alpha}}{(1+r^2/r_c^2)^{3\beta-\alpha/2}} + \frac{n_{02}^2}{(1+r^2/r_{c2}^2)^{3\beta_2}},
\end{equation}
where $n_e$ and $n_p$ are the electron and proton densities,
respectively. We allow all parameters to vary.

We model the temperature profile with a simple power-law: 
\begin{equation}
T_{\rm 3D}(r) = T_0  (r/r_t)^{-a}.
\end{equation}
This profile is then projected along the line of sight using the
formula of the spectroscopic-like temperature: 
\begin{equation}
T_{\rm los}=\frac{\int  WT_{\rm 3D} d V}{\int W  d V}, 
\end{equation}
where $W=(n_pn_e)/(T_{\rm  3D}^{0.75})$.
All the best fit parameters are determined using a $\chi^2$ minimization
technique applied to the models and the data. 
Finally, the 3D density and temperature profiles are used to estimate
the total gravitational mass through the equation of hydrostatic
equilibrium (HSE; \citealt{sarazin88}): 
\begin{equation}
M(r)=-3.68 \times 10^{13} T(r)r^2\left( \frac{d\log \rho_g}{d r} + \frac{d\log T}{d r} \right) {\rm h}_{70}^{-1}M_{\odot},
\label{eq:mhe}
\end{equation}
where the numerical factor includes the gravity
constant, proton mass, and the mean molecular weight,
$\mu=0.5954$. 

\begin{figure*}
\centering
\includegraphics[scale=0.4]{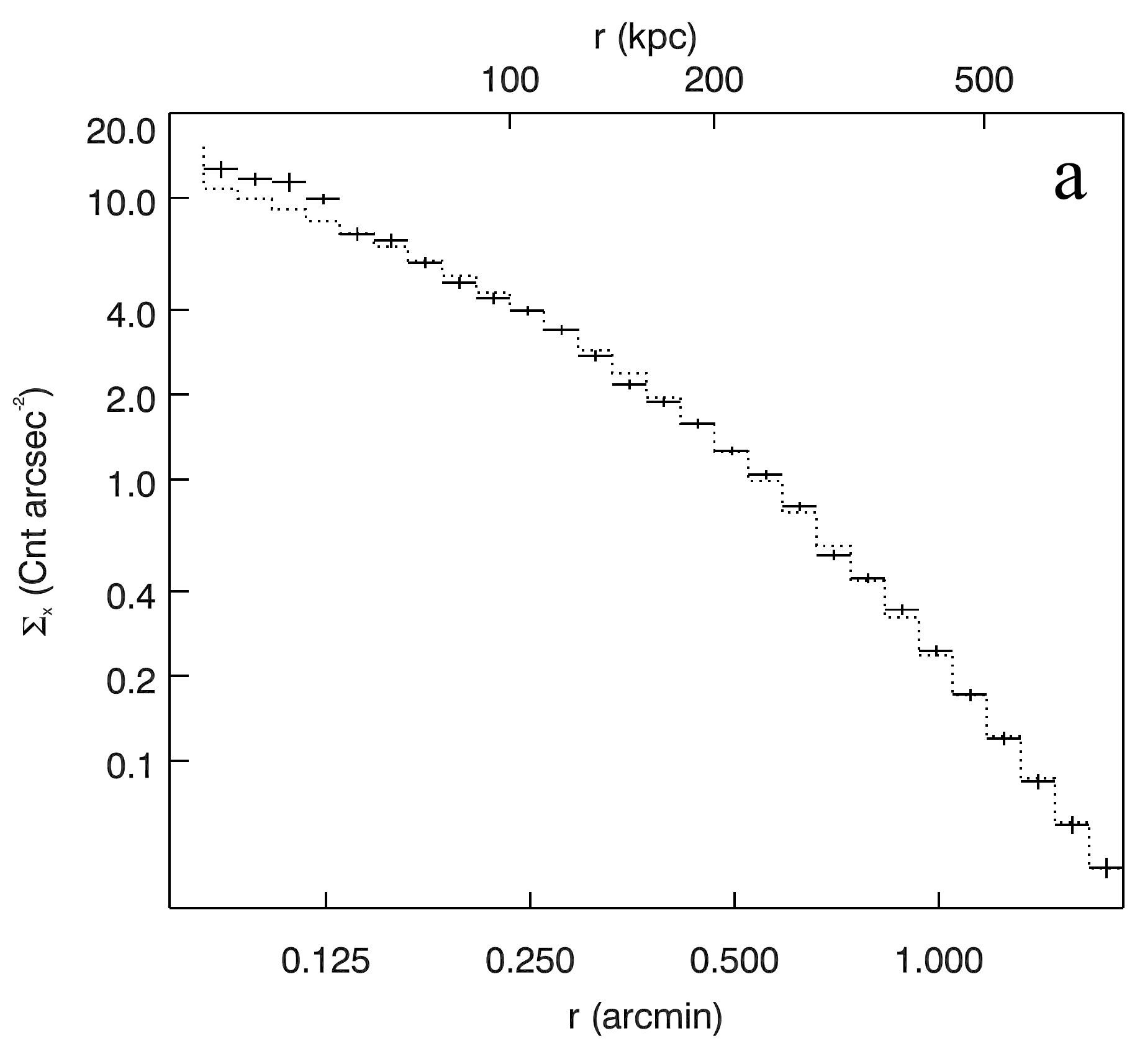} 
\includegraphics[scale=0.4]{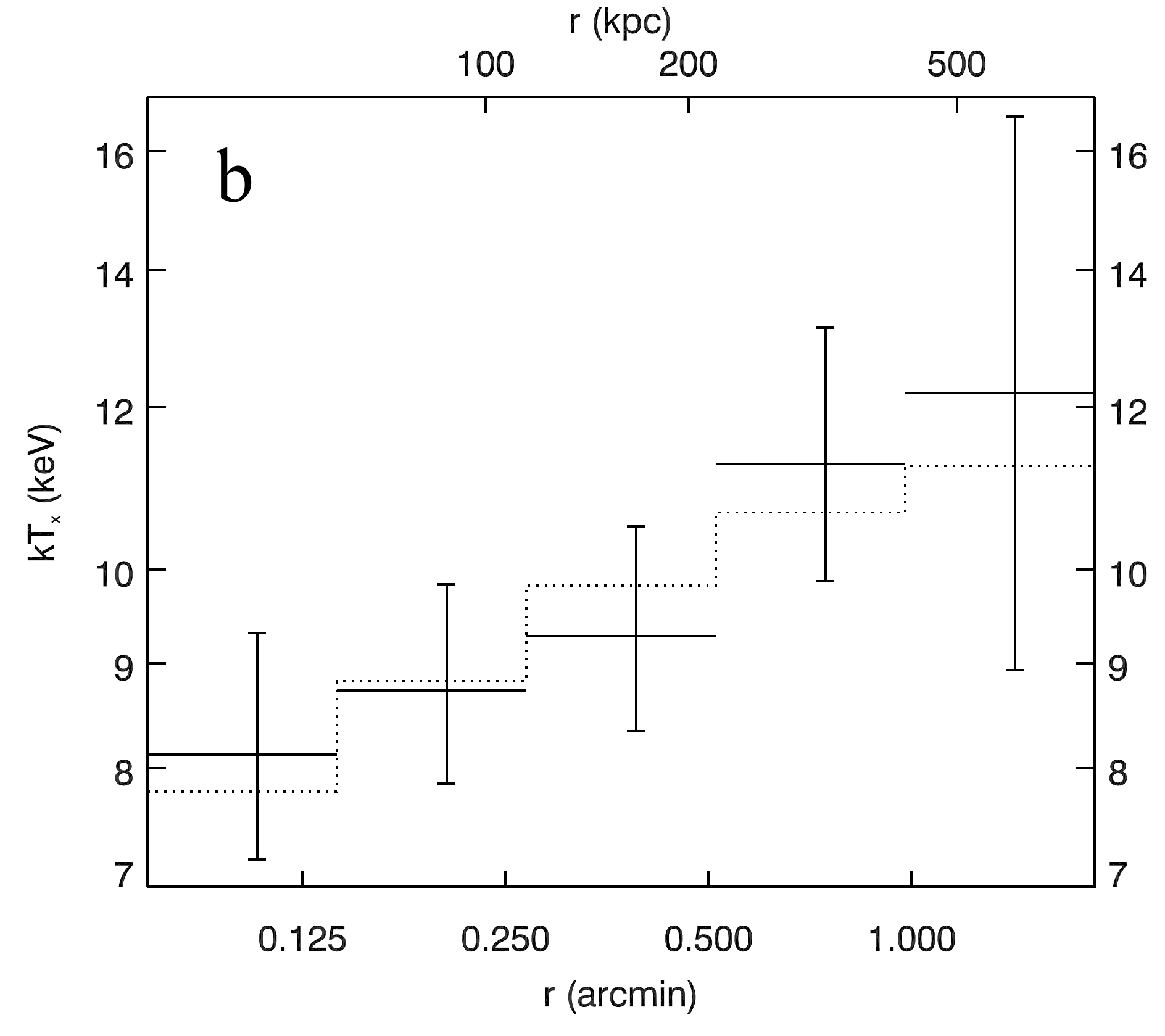} 
\includegraphics[scale=0.4]{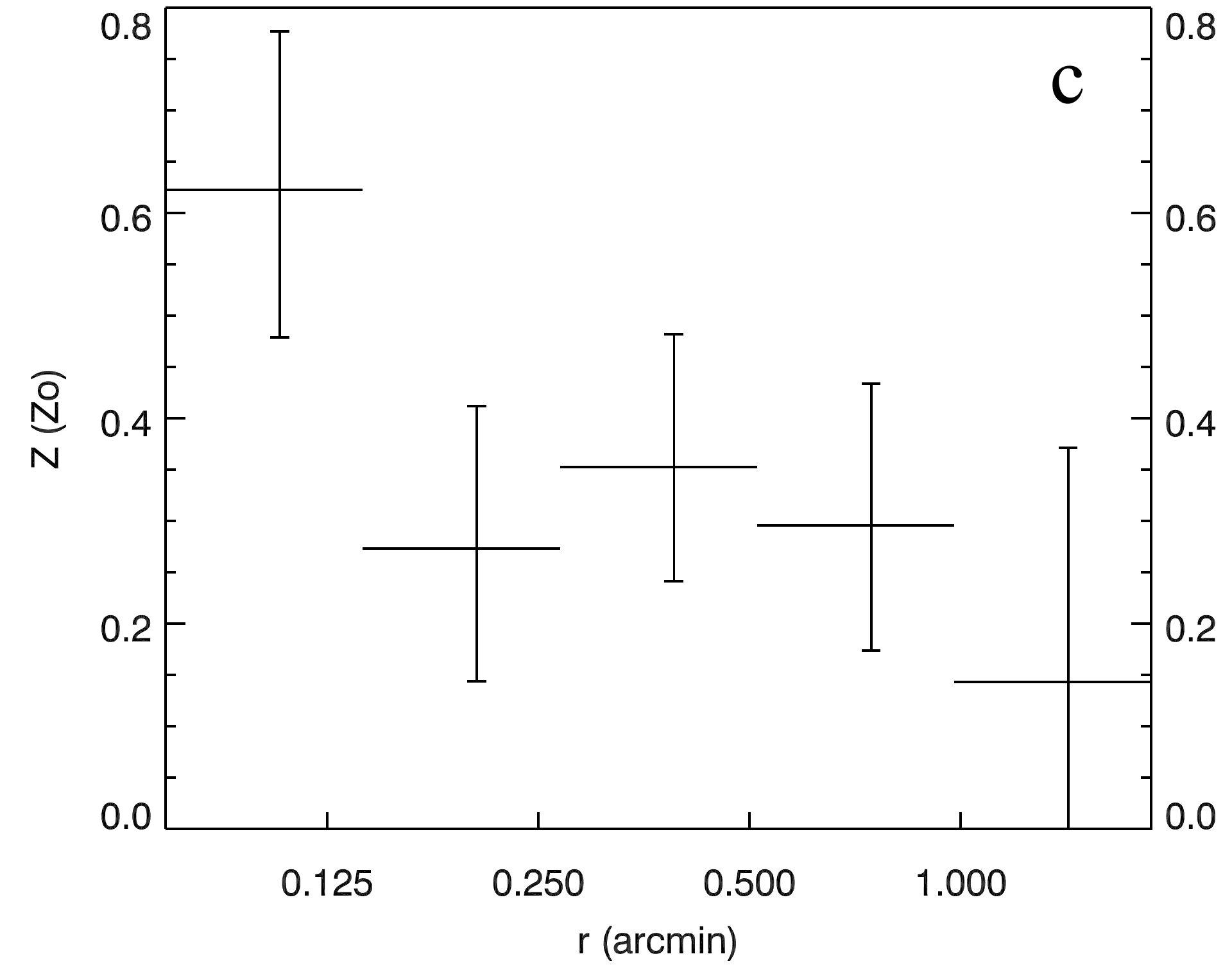} 
\includegraphics[scale=0.4]{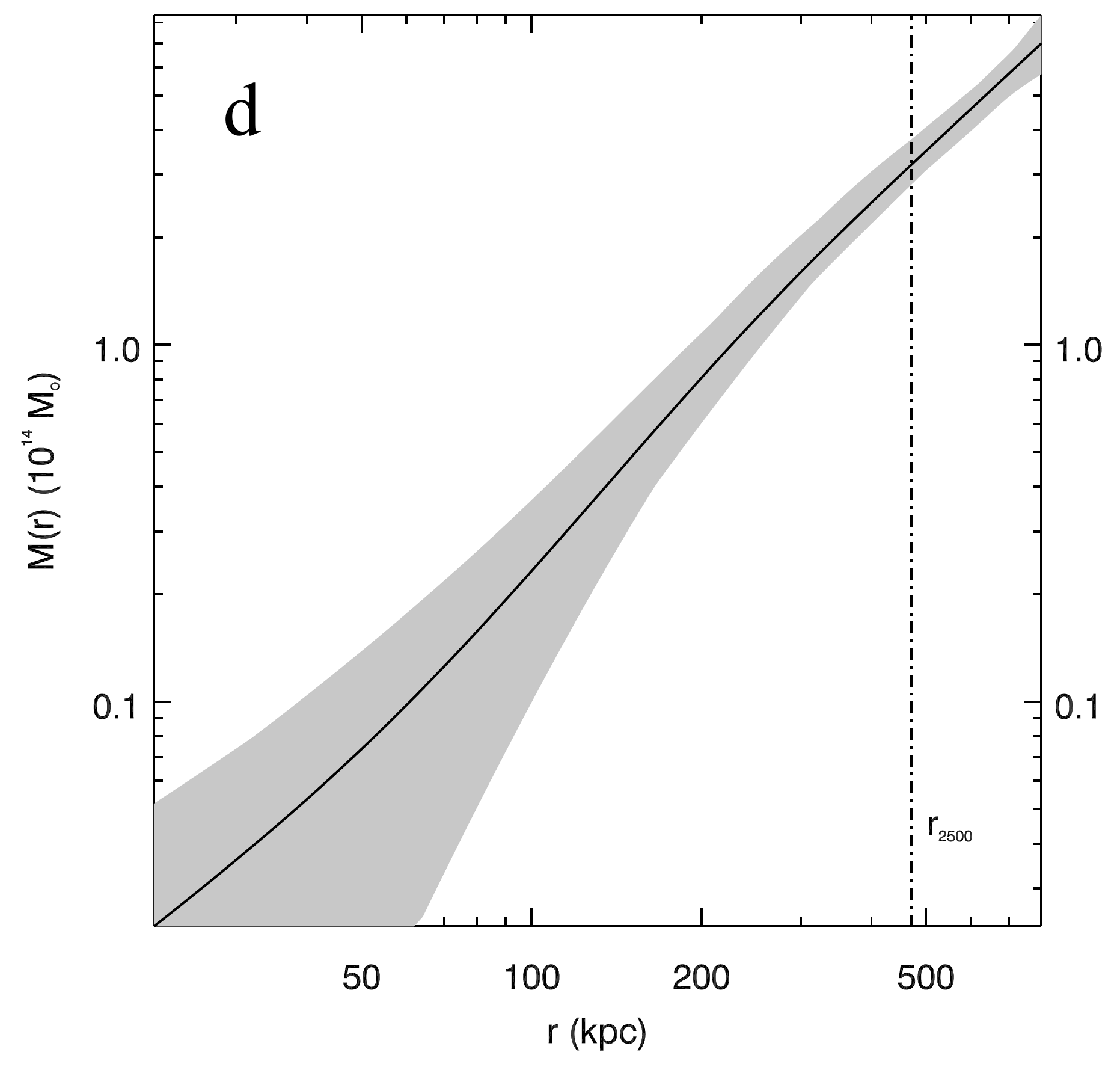} 
\caption{Radial profiles derived from the analysis of  {\it Chandra} Cycle 13
  150 ks X-ray observation, extended out to 800 kpc.
(a) {\it Top-left:} Radial profile of the X-ray surface brightness in the soft band
[0.5-2.5] keV (data points). The dotted line represents the best fit
model.  
(b)  {\it Top-right:} Projected temperature profile, clearly showing a temperature
  decrease towards the center of the cluster, indicating a cool core.
  The dotted line represents the best fit model.
(c)  {\it Bottom-left:} Metallicity profile 
(d)  {\it Bottom-right:} Hydrostatic mass profile (solid line) and its 1-$\sigma$
uncertainty (shaded area). The vertical line indicates $R_{2500}$.}
\label{xrayprof}
\end{figure*}

To estimate the uncertainties, we produce 500
realizations of the surface brightness and temperature profiles
assuming a Poisson distribution for the total counts in each annulus
and a Gaussian distribution for the projected temperature. The fitting
procedure described above is repeated each time. The derived mass
profiles are related to the original HSE mass profile via the resulting
value of the least-mean-square formula ($=\Sigma [M_{\rm realization}
- M_{\rm original}]^2$, where the sum extends to all radial bins). We
consider the 68\% of the profiles (340 in number) with the smallest
associated value and, finally, for each radial bin, we consider the
maximum and minimum values of the selected profiles. 
The resulting mass profile and its uncertainties are plotted in Figure~\ref{xrayprof}d.
The HSE radius and
mass at overdensity $\Delta=$2500 are  
$R_{2500} =   471^{+ 54}_{-33}$\h~ kpc, and 
$M_{2500} =   3.2 ^{+ 0.6}_{-0.3}\times10^{14}$\h\msun, respectively, 
and the gas mass within this radius is
$M_{\rm gas, 2500}= 4.4^{ +0.8}_{-0.5} \times 10^{13}$\h\msun. 

On average, HSE masses are expected to be biased low by
10-15\% as evident from simulations \citep{rasia06, nag07, battaglia13}
and observations \citep{mahdavi08, mahdavi13}. This offset
is smaller than our statistical uncertainty. 

The value of $R_{2500}$ is within the region probed
by the observation, and thus its measurement is conservative and
robust. 
However, the lower overdensities $\Delta=500$ and $\Delta=200$ are not
within the observed region, and we therefore need to
extrapolate. For that purpose, we follow two different approaches. 
 \paragraph{1: NFW-mass extrapolation} We fit the 500 realizations
with the Navarro-Frenk-White (NFW, \citealt{nfw95, nfw96, nfw97})
formula: 
\begin{equation}\label{eq.nfwmass}
M_{\rm NFW}=4 \pi M_0 r_s^3 \left[ \log(1+r/r_s) -\frac{r/r_s}{1+(r/r_s)}  \right],
\end{equation}
where $r_s$ is the scale radius and $M_0$ the normalization of the
mass profile. The fitting is carried out only within the observed
radial region. The results lead to
$R_{500}=1.15^{+0.59}_{-0.25}$\h~ Mpc and
$R_{200}=1.78^{+1.24}_{-0.43}$\h~ Mpc. The errors 
represent the minimum and maximum values of the 68\% of the
analytic expressions that are the closest to the NFW fit of the
original mass profile. The resulting extrapolated masses are  
$M_{500}=1.1^{+ 0.9}_{-0.6} \times 10^{15}$\h\msun~and  
$M_{200}=1.8^{+ 1.8}_{-0.7} \times 10^{15}$\h\msun.
\paragraph{2: $M-Y_X$ relation.} To derive $R_{500}$ we also apply the
iterative method based on the $M-Y_X (=M_{\rm gas} \times T_X) $
relation proposed by \cite{kravtsov.etal.06}. We start with an initial
guess for the radius (we consider twice the value of the measured
$R_{2500}$). We evaluate the gas mass at that radius from the surface
brightness profile and compute the X-ray temperature from the spectra
extracted in the spherical shell with maximum and minimum radii equal
to  the specific radius and 15\% its value. The obtained $Y_X$ is
compared with the $Y_X-M$ relation calibrated from hydrostatic mass
estimates in a nearby sample of clusters observed with {\it Chandra}
\citep{vikhlinin.etal.09}. This returns an estimate of $M_{500}$ and,
thus, a new value for $R_{500}$. The process is repeated until convergence
in the radius estimate is reached. The resulting radius is 
$R_{500}=1.27^{+0.10}_{-0.08}$\h~Mpc, corresponding to 
$M_{500}  = 1.27^{+ 0.31}_{-0.22} \times 10^{15}$\h\msun. 

While the two extrapolation methods agree within errors, we note that the
NFW-mass extrapolation results in much larger uncertainty.  This is
due to the fact that the HSE mass profile is constrained at a small
radius (just above \rtfh) and thus the external slope of the
cluster is poorly constrained from X-ray observations.

\subsection{Sunyaev Zel'dovich Array Observations and SZ  Mass Estimates}\label{s.sz}
\begin{figure}
\centering
\includegraphics[scale=0.25]{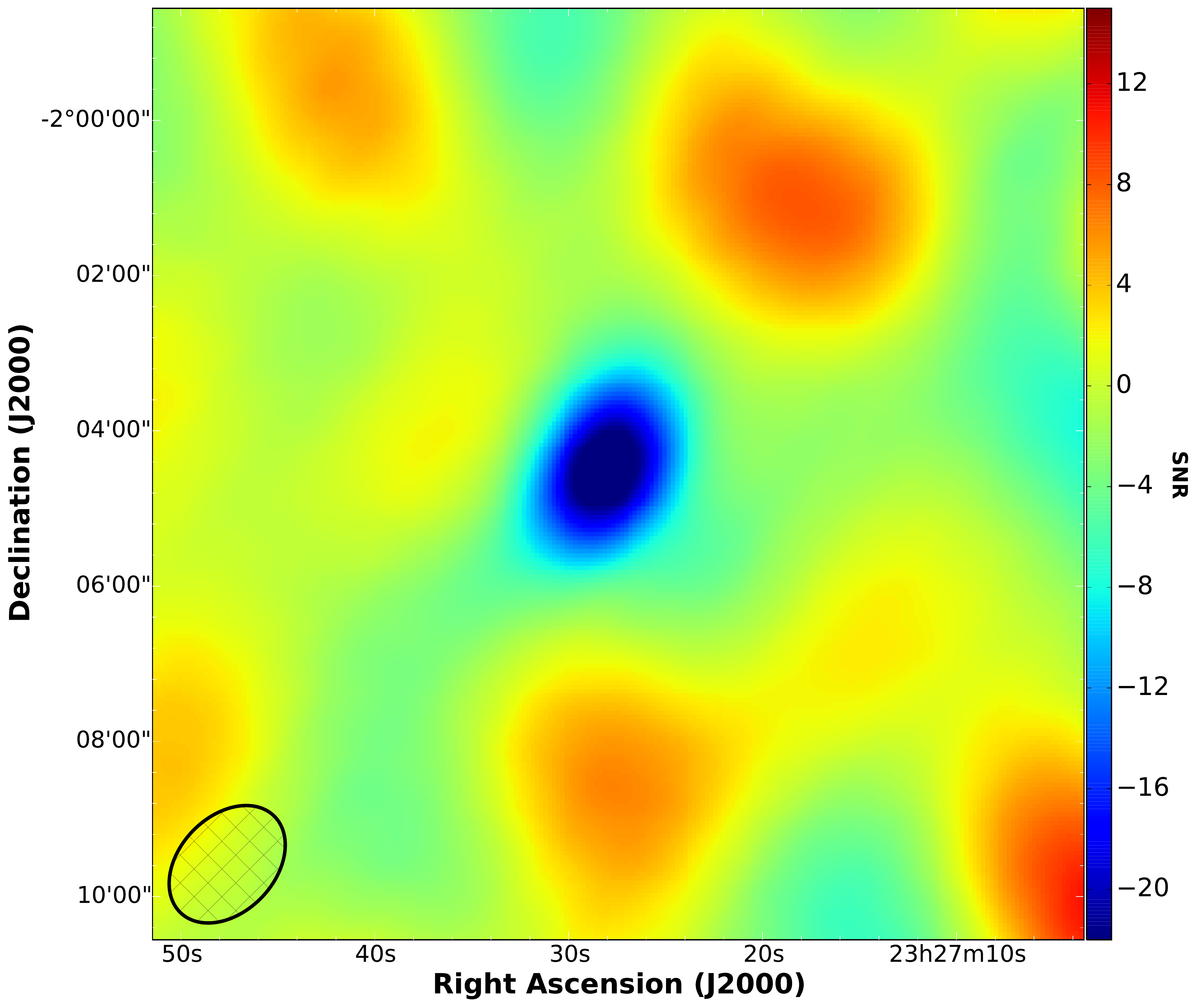}
\caption{SZ observation of \clustername. The CLEANed image is
$8\arcmin \times 8\arcmin$ in size. The image is in units of SNR and made
from the SZA $uv$-data with a Gaussian $uv$-taper of 0.1 at 4$k\lambda$. The
FWHM of the synthesized beam for the image is shown in the bottom left
corner. In this image, SNR of 1 corresponds to { 0.21}$~\rm{mJy/bm}$.}
\label {fig:sz}
\end{figure}

The Sunyaev Zel'dovich Array (SZA) observed RCS2327 for a total of
 48
hours between 2007 September and November. The SZA is an eight-element
interferometer with 30 and 90~GHz receivers. The SZA was configured in a
standard configuration with 6 telescopes arranged in a compact, short-baseline
configuration with two outlying telescopes $\sim$30~m from the central group.
The correlated bandwidth was 8 GHz, centered on 31 GHz, resulting in
projected lengths of 350$-$1300$\lambda$ on the short (SZ-sensitive)
baselines and 2000$-$8000$\lambda$ on the longer baselines. The data were
calibrated and flagged using the MATLAB pipeline
described in \citet{muc07}; { 41\%} of the data were removed, largely due
to shadowing in the compact array, which is increased in equatorial
and lower declination objects. 
The rms noise level
in the short-baseline data is
{ 0.21}~mJy~beam$^{-1}$, corresponding to a { 15}~$\mu$K rms brightness
temperature in the 1.9\arcmin$\times${ 2.8}\arcmin\ synthesized beam. A
bright radio source is detected nearly 9\arcmin\ to the west
of the cluster, but it does not affect the SZ
detection.
{ There is one faint radio source coincident with the
cluster which we jointly model when presenting data. This source is present
in the NRAO VLA Sky Survey (NVSS, Condon 1998).}
The deconvolved image of the cluster, after subtraction of
the radio { sources}, is shown in Figure~\ref{fig:sz}. The peak
significance in this image is { 22}$\sigma$.

The SZA interferometer acts as a spatial filter sensitive to the
Fourier transform of sky emission on angular scales determined by the
baseline lengths. To recover the integrated Compton-$y$ parameter, $Y$,
we fit our Fourier plane data to the transform of the generalized NFW
pressure profile presented in \citet{nag07}, which was motivated by
simulations and X-ray
observations \citep{mro09}. In this five-parameter model we fix the three shape
parameters ($\alpha$, $\beta$, $\gamma$) to the best fit values
derived from X-ray observations of clusters \citep[1.0620, 5.4807, 1.156;][]{arn10}.
We fit the profile normalization and scale radius.
The cluster centroid and the { flux of one emissive source} are
allowed to vary as well. The models are fit to the data directly in the
$uv$-plane, which correctly accounts for the noise in the data.

To determine { the significance of the cluster detection}, we
compare the $\chi^2$ of the best fit model 
including the cluster and emissive source with the $\chi^2$ of the best fit
model including only the emissive source. Expressed in terms of Gaussian
standard deviations, the significance of the SZ detection is { 30.2}$\sigma$.

\subsection{Estimates of the Y parameter}\label{s.szy}
{ We compute two estimates of the $Y$ parameter, $Y_{sph}$ and $Y_{cyl}$.
$Y_{sph}$ is a spherical integral of the pressure profile. 
It is relatively insensitive to unconstrained modes in the interferometer-filtered data
and proportional to the total integrated pressure of the cluster, making
it a robust observable (see Marrone et al 2011). }

{ To compute $Y_{sph}$,} we volume-integrate the radial profile to an overdensity radius, $r_\Delta$,
\begin{equation}
Y_{\Delta,sph} = \frac{\sigma_T}{m_e c^2} \int_0^{r_\Delta} P(r/r_s) dV,
\end{equation}
as in \citet{mar12}.
We determine the overdensity radius of integration by enforcing consistency
with the $Y_{500,sph} - M_{500}$ scaling relation derived by \citet{and11}.
To enforce consistency, we iteratively chose the integration radius (and, by
extension, the mass) until the mass and $Y_{\Delta}$ lie on the mean relation.
{ This analysis yields
$Y_{500,sph} = 13.4 \pm1.0 \times 10^{-5} ~\rm{Mpc^2}$ with
$R_{500} = 1.13 \pm 0.02~\rm{Mpc}$ (this radius is 3-15\%
smaller than the $R_{500}$ that we derive from extrapolating the X-ray
data in Section~\ref{s.xray}). }

{ $Y_{cyl}$ is a 
cylindrical integral of the pressure profile along the line of sight,
\begin{equation}
Y_{\Delta,cyl} = \frac{\sigma_T}{m_e c^2} \int_0^{r_\Delta}   d\Omega \int_{-\infty}^\infty P_e dl
\end{equation}
\citep[see also][]{mro09}. $Y_{cyl}$ corresponds to the aperture
integrated SZ flux, and is sensitive to the line of sight contribution
of pressure beyond the radius of interest.  For our gNFW fits, the
ratio of $Y_{cyl}$/$Y_{sph}$ in the $211\farcs8$ aperture is $1.0564$. 
We compute $Y_{500,cyl}$ to directly compare our results with the
SZ observations of \clustername~ reported by \citet{hasselfield13}
with data from the Atacama Cosmology Telescope (ACT). 
From their ``Universal Pressure Profile'' (UPP) analysis, which
implicitly imposes the $Y-M$ scaling relation of Arnaud (2010),
they obtain $Y_{500,cyl} = 19.1 \pm0.2 \times 10^{-5} ~\rm{Mpc^2}$
 within an aperture of $R_{500}=2.8\pm0.1$ arcmin ($1.22\pm0.04$ Mpc at the cluster
redshift). Using the same aperture, we measure $Y_{500,cyl} =
16.8^{+1.6}_{-1.4} \times 10^{-5} ~\rm{Mpc^2}$, within $1.5 \sigma$ of
  the \citet{hasselfield13} measurement. }

\subsection{SZ Mass Estimates}\label{s.szy}
{ We estimate the cluster mass from the value of $Y_{sph}$ quoted
  above, which corresponds to 
$M_{500} = 8.5\pm1.1\times10^{14}$\h\msun~ using the \citet{and11}
scaling relation.  The uncertainty assumes 21\%  scatter in $Y$ at
fixed mass in the \citet{and11} scaling relation \citep{Buddendiek14}.} 


{ 
We also estimate the cluster mass by applying the method outlined in
Mroczkowski (2011, 2012) to the SZA data.  This method assumes the gas
is virialized and in thermal hydrostatic equilibrium within the
cluster gravitational potential.  Further, we assume the total matter
density $\rho_{\rm  tot}$ follows an NFW dark matter profile \citep{nfw95}, with the gas density $\rho_{\rm  gas}$ is a
constant fraction of the total density ($\rho_{\rm  gas} = f_{\rm gas}
\rho_{\rm  tot}$), and the pressure and density profiles are
spherically symmetric.  This method has been applied 
successfully and compared with other mass estimates in several works
\citep[e.g.,][]{reese12, umetsu12, med13}.
A fit to a gNFW profile described above 
yields $R_{2500} = 0.54 \pm 0.01~\rm{Mpc}$, $M_{2500} = 4.6\pm 0.5
  \times10^{14}$\h\msun, assuming an average gas fraction within $R_{2500}$ of $f_{gas}=0.137$ from the
x-ray analysis in \S~\ref{s.xray}. 
This method yields $R_{500} = 1.15 \pm 0.04~\rm{Mpc}$, $M_{500} =
8.9\pm 0.9 \times10^{14}$\h\msun, assuming an average gas fraction in this
radius of $f_{gas}=0.12$ from Menanteau et al. (2012). 
We estimate a $\sim10\%$
scatter due to the assumption on average gas fraction value and other
model assumptions.

Our mass estimates are consistent with \citet{hasselfield13}, who measure $M_{500}=9.4\pm1.5
\times10^{14}$\h\msun~ from the UPP Y parameter quoted above.
In addition to the UPP mass, \citet{hasselfield13}
report a range of higher $M_{500}$ estimates based on different scaling relations,
$M_{500} = 12.5-14.3 \times10^{14}$\h\msun, somewhat higher than our
measurement. 
However, the inconsistency
between the higher-mass \citet{hasselfield13}
values and our measurement is not significantly worse than the inconsistency
with their own UPP mass.}


\subsection{Wide Field Imaging and Weak-Lensing  Mass Estimates}\label{s.wl}
We obtained deep wide-field imaging data for RCS2327 in the $i'$
filter using Megacam on CFHT with the aim of determining the mass
using weak gravitational lensing. The observing strategy and weak
lensing analysis follows that of the Canadian Cluster Comparison
Project \citep[CCCP;][]{hoekstra12}, with the only difference
that we use the $i'$ for the weak lensing analysis. The $i'$ data
consist of 8 exposures of 650 s each, which are combined into two sets
(each with a total integration time of 2600 s). The pointings in each
set are taken with small offsets, such that we can analyse the data on
a chip-by-chip basis. 

{ The various steps in the analysis, from object detection to unbiased shape measurements 
and cluster mass, are described in detail in \cite{hoe07},  with updated procedures in \cite{hoe15} and 
we refer the reader to those papers for more details. We measure galaxy shapes as described in \cite{hoe15}, which
includes a correction for multiplicative bias based on simulated images. The
resulting shapes are estimated to be accurate to $1-2\%$, much smaller
than our statistical uncertainties.} The shape measurements for each set of exposures are then combined
into a master catalog which is used to derive the weak lensing
mass. To reduce contamination by cluster members, we also obtained
four 720 s exposures in $r'$, which are combined into a single
image. Galaxies that are located on the cluster red-sequence are
removed from the object catalog, which reduces the level of
contamination by a factor of two. However, many faint cluster members
are blue, and we correct the lensing signal for this residual
contamination, as described in \cite{hoe07}. 

\begin{figure}
\centering
\includegraphics[scale=0.45]{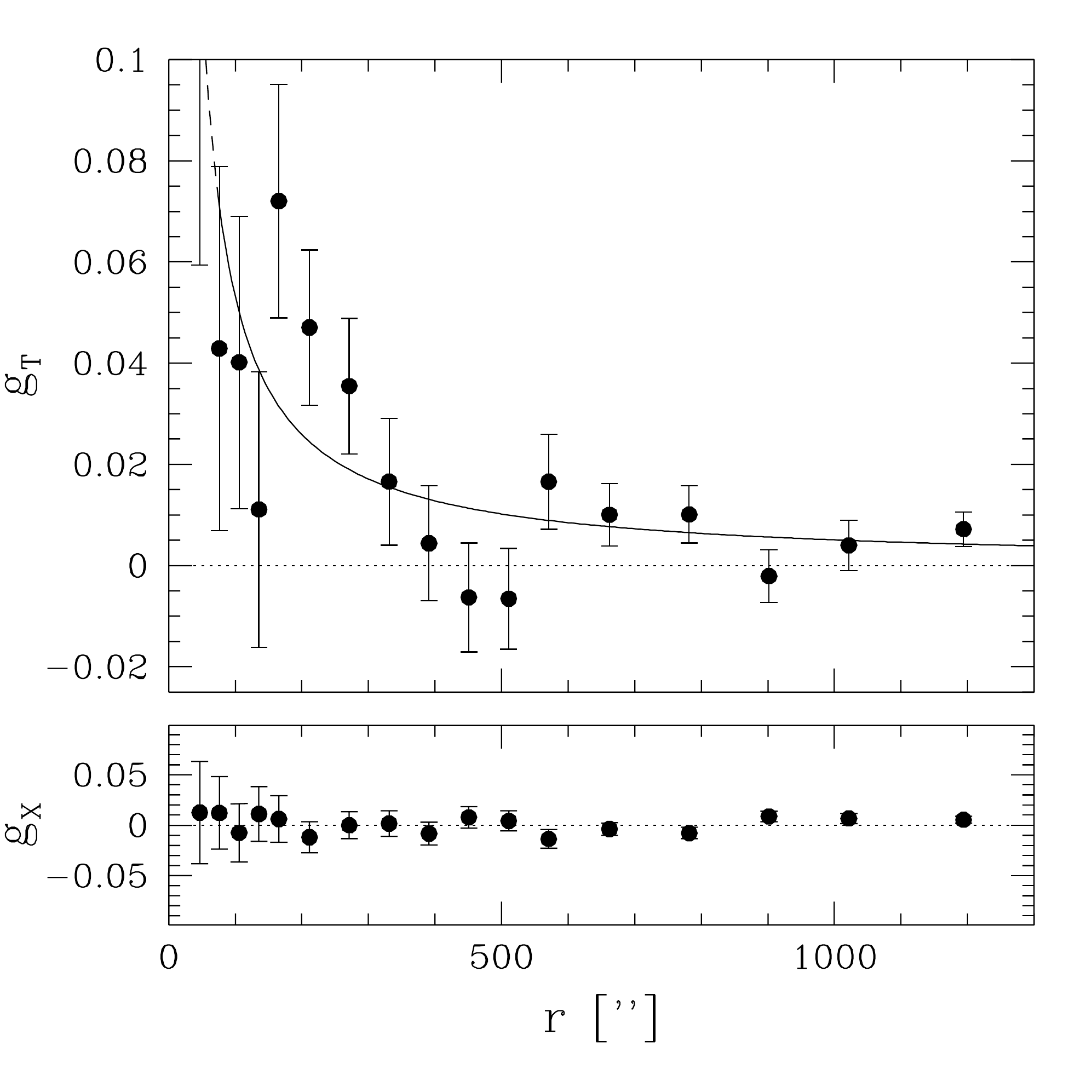} 
\caption{Mean tangential shear as a
function of radius from the BCG is shown in the top panel, along with
the best fitting isothermal sphere model for reference. { The model was
fit only to the data at $r > 75''$ (solid line). The dashed line is the
extrapolation of the model to smaller projected radii.  } The bottom
panels shows the 'B'-mode lensing signal, which should be consistent
with zero if systematic distortions have been correctly removed.}
\label{wl}
\end{figure}

To quantify the lensing signal, we compute the mean tangential shear
as a function of distance from the cluster center using galaxies with
$22<i'<24.5$. Figure~\ref{wl} shows the resulting signal, which
indicates that the cluster is clearly detected. The bottom panel shows
a measure of the lensing `B'-mode, which is consistent with zero,
indicating that the various corrections for systematic distortions
have been properly applied.

As discussed in \cite{hoe07}, the weak lensing mass can be derived in a
number of ways. However, to relate the lensing signal to a physical
mass requires knowledge of the redshift distribution of the galaxies
used in the lensing analysis. We use the results from \cite{hoe15} and
find that the mean ratio of angular distances between lens-source and
observer-source is $D_{ls}/D_s=0.194$.

{ For reference, we show the best fit singular isothermal sphere model
in Figure~\ref{wl}, for which we obtain an Einstein radius
${\rm R}_{\rm E}=10\farcs1\pm 1\farcs9$, which yields a velocity 
dispersion of $\sigma=1345^{+122}_{-134}$ km s$^{-1}$ for the 
adopted source redshift distribution. This value is in excellent agreement
with the dynamics inferred from the galaxy redshifts. We also fit an
NFW model to the data, adopting the mass-concentration relation
suggested by \cite{duf08}, which yields a mass M$_{\rm
200}=2.0^{+0.9}_{-0.8}\times 10^{15}$\h\msun.} 
We compare the weak lensing mass to other mass estimates and in other
radii in \S~\ref{s.massproxycomp}.

\subsection{Strong Lensing  Mass Estimates}\label{s.sl}
\begin{figure*}
\centering
\includegraphics[scale=1]{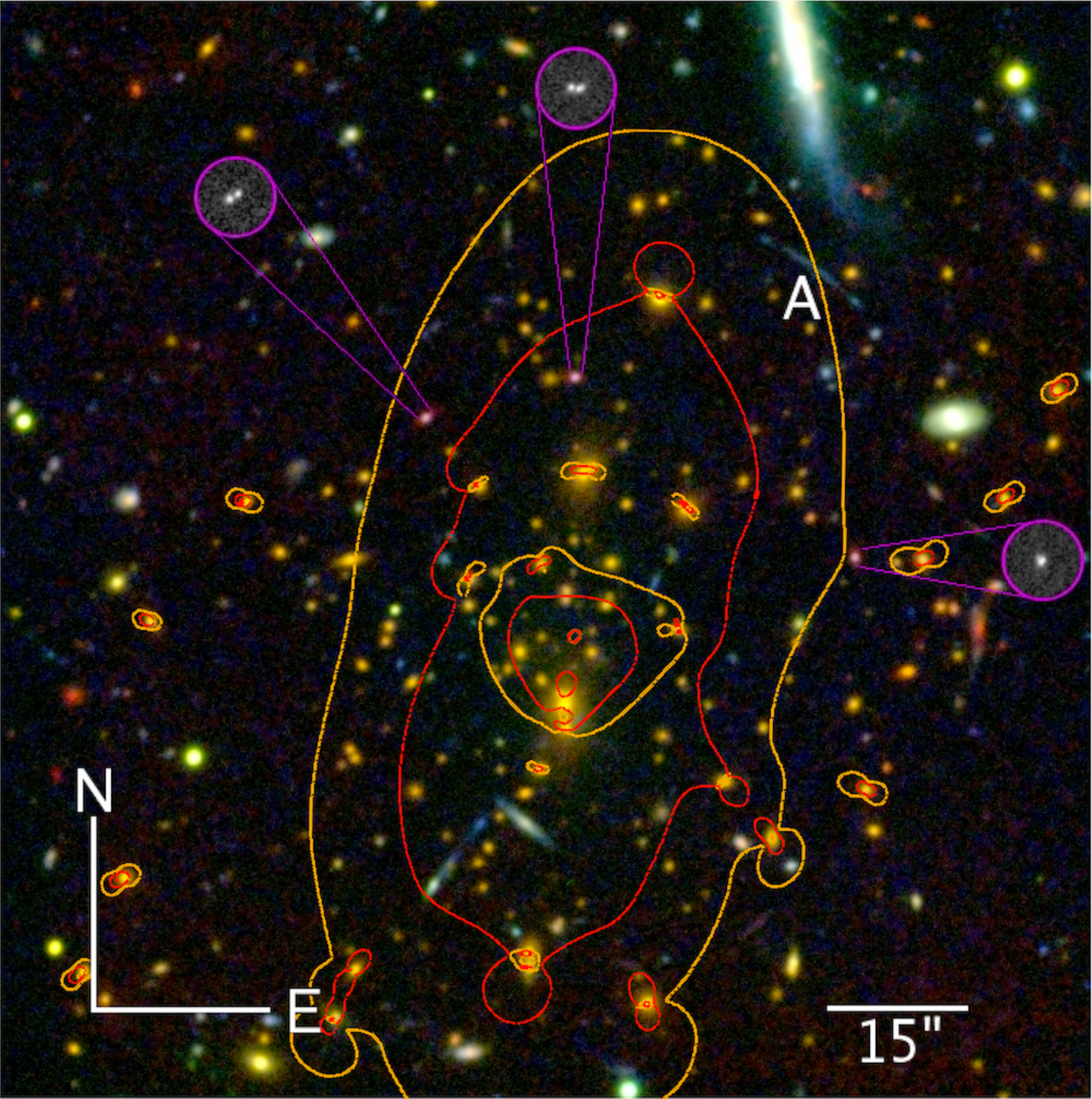} 
\caption{ A 2'$\times$2' color composite
image of the core of RCS2327 composed of images from \HST~ in the F435W
filter (blue channel), and from Magellan in the $i-$band (green
channel) and $K_s$-band (red channel). The point spread functions have
been matched to the worst image; the effective resolution is
$\sim0\farcs6$. The merging pair image of source A at $z$=2.9834 is
indicated. The greyscale cutouts are 2\arcsec~~ in diameter and show
the full-resolution images of source B, at $z$=1.4155, in the F435W
filter. The critical curves from the best fit lens model are overplotted, in
red for a source at $z$=1.415, and in orange for a source at $z$=2.9834. }
\label{irim}
\end{figure*}

RCS2327 was observed by \hst+ACS (Cycle 15 program GO-10846; PI Gladders) as part of a larger
effort using both ACS and NICMOS to acquire deep multi-band imaging of
this cluster. Unfortunately, the failure of ACS in early 2007
truncated this program, and the only complete image which was acquired
is a 3-orbit F435W image of the cluster core taken using the ACS Wide
Field Channel\footnote{The field of \clustername~ was recently imaged by \hst~ in
  Cycle 20 program GO~13177 (PI Brada{\v c}). These data are not used in
  this paper. A forthcoming lensing analysis of the Cycle 20 data will
be presented in \citet{hoag15}.}. 
Additional available observations of the cluster core
relevant for the strong lensing analysis include a deep ($\sim$2
hours) K-band image of RCS2327 acquired using the PANIC instrument on the
Baade Magellan I telescope in 2006, as well as an incomplete
4-pointing mosaic of RCS2327 in the F160W filter taken with \HST+NICMOS;
we have reconstructed this last image from the useable portions of a
nominally failed \HST~ observation which nevertheless yielded some
useful frames in a single orbit before guiding issues truncated the
remainder of the observations. A color composite image of the cluster
core, made from the F435W image, the deep LDSS-3 $i$-band image (see
\S~\ref{s.richness} above), and the PANIC K$_s$-band image, is shown in Figure \ref{irim}.

Using these various imaging data, we identify two sets of
multiply-imaged galaxies that are lensed by RCS2327 for which we have
acquired spectroscopic redshifts as part of the overall spectroscopic
program described in \S~\ref{s.spec}. Both sources are indicated in
Figure~\ref{irim}. A merging pair of images of source A is located at 
23:27:29.43, $-02$:03:47.8, to the NE of the cluster center. Its redshift,
$z=2.9834\pm 0.0010$, is determined from a strong Ly$\alpha$ emission
line present in the early LDSS-3 spectroscopy described above.  This
lensed source was apparent in the RCS2 discovery imaging data, with
 a remarkably large separation from the cluster center, 
R=56\farcs8, as
measured from the BCG. The arc does not appear to
be caused by local substructure in the cluster, as there are no nearby
significant cluster galaxies. 

Source B was observed spectroscopically in queue mode in semester
2007B using the Gemini South telescope with the GMOS instrument. We
observed RCS2327 for $8\times1800$s in multi-object spectroscopy
mode. The observations were taken with the B600\_G2353 grating, no
filter, and the detector binned 1$\times$2 (spatial$\times$spectral
axes), resulting in wavelength coverage of $\sim 2700$\AA~per slit,
and a spectral resolution of $\sim240$ km~sec$^{-1}$.  The grating tilt
was optimized to record a wavelength range of $\sim3800-6500$\AA~~for
images of source B.

The redshift of source B is $z=1.4155\pm 0.0008$, based on
\OII~ emission present in a GISMO observation (see \S~\ref{s.spec})
confirmed by several FeII lines in absorption in the Gemini spectra.
Source B is lensed into three images, and is not morphologically
obvious in the discovery data from RCS2327, as it is not lensed into a
classic tangential arc. It is apparent in the combined \HST~ and IR
imaging since it has a unique color and internal morphology. These
properties also allow us to robustly eliminate the presence of a
fourth counter image; source B is lensed as a naked cusp configuration
\citep[e.g.,][]{ogu04}.  A close inspection of the F435W image reveals
that two of the images (B1 and B2) have two emission knots at their
center; overall source B appears to be a compact galaxy with a
primarily redder stellar population, but with two well confined
regions of active star formation in the galaxy's core. The detailed
position of these bright knots indicates a larger magnification in the
tangential direction than in the radial direction for this source. The
two knots in the third image are not resolved, but the image is elongated
in the tangential direction. Source B also has a significant Einstein
radius, with separations for the three images from the cluster BCG of
36\farcs8, 36\farcs6, and 35\farcs8. Further lensed features are also
apparent, but we do not yet have redshift information for
them and they are not used in the initial
lensing model discussed below.
     
A strong lensing model for RCS2327 was constructed using the publicly
available software LENSTOOL \citep{jul07}. The mass model is composed of multiple mass
clumps. The cluster halo is represented by a generalized NFW
distribution \citep{nfw97}, parametrized with position, $x$, $y$;
ellipticity $e$; position angle $\theta$; central slope $\alpha$; and
concentration $c$.
The 50 brightest red-sequence-selected cluster-member galaxies
are represented by Pseudo-Isothermal Ellipsoidal Mass Distributions
\citep[PIEMD; see ][for details]{jul07} parametrized with positional parameters ($x$, $y$, $e$, $\theta$) 
that follow their observed measurements, $r_{\rm{core}}$ fixed at 0.15 pc, and 
$r_{\rm{cut}}$ and $\sigma_0$ scaled with their luminosity (see Limousin et al.\ 2005 
for a description of the scaling relations). The parameters of an L*
galaxies were fixed at $r_{cut}=40$ kpc and $\sigma_0=160$ km s$^{-1}$.   
The model consists of 13
free parameters. All the parameters of the cluster halo are allowed to
vary (R.A., Decl. of the mass clump, ellipticity, position angle,
scale radius, concentration and central radial mass profile).

The
constraints are the positions of the lensed features and their
redshifts. Each component of Arc A was represented by
three positions, and the two cores of source B were used in each of
its images. The best fit model is determined through Monte Carlo
Markov Chain (MCMC) analysis through minimization in the source plane,
with a resulting image-plane RMS of 0\farcs17. The best fit
parameters and their 68\% percentile
uncertainties are presented in Table~\ref{table.lensmodel}. Some of the
model parameters are not well-constrained by the lensing evidence. In 
particular, a large range of values is allowed for $r_s$ and $\alpha$,
and the model can converge on any value of the concentration parameter
$c$. The latter is not surprising, since in order to determine the
concentration parameter one needs to constrain the slope of the mass
profile on small and large radii, beyond the range of the strong
lensing constraints. Thus the concentration uncertainty
given in Table~\ref{table.lensmodel} represents the range of priors assumed
in the lens modeling process.
We find strong correlations between, $\alpha$, $r_s$, and $c$, which
we fit to find \Pasfit~ and \Pacfit.  

\begin{deluxetable*}{lllllllll} 
 \tablecolumns{8} 
\tablecaption{Best fit Strong Lensing Model Parameters  \label{table.lensmodel}} 
\tablehead{\colhead{Halo }   & 
            \colhead{Model}     & 
            \colhead{RA}     & 
            \colhead{Dec}    & 
            \colhead{$e$}    & 
            \colhead{$\theta$}       & 
            \colhead{$r_{\rm s}$} &  
            \colhead{$\alpha$}  &  
            \colhead{$c$}\\ 
            \colhead{}   & 
            \colhead{}   & 
            \colhead{($\arcsec$)}     & 
            \colhead{($\arcsec$)}     & 
            \colhead{}    & 
            \colhead{(deg)}       & 
            \colhead{(kpc)} &  
            \colhead{}  &  
            \colhead{}  } 
\startdata 
Halo 1   & gNFW  & \Px        & \Py           & \Pe       & \Ptheta        &\Prs     & \Palpha    & \Pc \\ 
\enddata 
 \tablecomments{Coordinates are measured in arcseconds East and
   North of the center of the BCG, at [RA, Dec]=[351.865026, $-2$.076924]. The
   ellipticity is expressed as $e=(a^2-b^2)/(a^2+b^2)$. $\theta$ is
   measured North of West. Error bars correspond to 1-$\sigma$
   confidence level as inferred from the MCMC optimization. 
}
\end{deluxetable*} 

The Einstein radius of a lens is often used as a measure of its
lensing cross section, or strength. We measure the effective Einstein
radius as $R_E=\sqrt{A/\pi}$, where $A$ is the
area enclosed by the tangential critical curve, 
$R_E(z=1.4155)=25\farcs9$ for source B, and  $R_E
(z=2.9834)=40\farcs2$ for the giant arc A. These radii
are smaller than the separations between the arcs and the BCG, due to
the ellipticity of the lensing potential. 
The mass that results from the lensing model can be quoted at a range
of radii, though it is clear that the mass is most robustly measured
at the critical radii probed by the lensed images used to construct
the model \citep[e.g.][]{meneghetti10}. We integrate the strong lens model
within circular apertures at radii corresponding to the mean positions
of sources A and B with respect to the position of the main cluster
NFW halo, and { find enclosed projected mass
  $M_{cyl}(<R_A)=$\MarcAm,  $M_{cyl}(<R_B)=$\MarcBm. 
Statistical} uncertainties are computed by sampling models described
by the MCMC outputs, considering only models with values of $\chi^2$
within two of the best fit, representing 1-$\sigma$ uncertainty in the
parameter space. The resulting masses are a
measure of the projected (i.e., cylindrical) masses within the quoted
radii. These statistical uncertainties may fail to reflect some
systematics due to the small number of lensing constraints in this
system. In particular, since the lens is only constrained by arcs on
one side of the cluster, we see correlations in the parameter space between the mass,
ellipticity, and position of the lens. The superior data expected
from \hst~ Cycle~20 program GO-13177, will enable a better constrained
lens model \citep{hoag15}. { We adopt a 15\% systematic uncertainty
  from Zitrin et al. (2015) for clusters with similar strong lensing signal.} 


A notable further result from the strong lens model is that the
cluster halo is offset from the BCG by \Px~ and
\Py~ arcseconds in right ascension and declination,
respectively. This corresponds to an offset of 54 kpc at the
cluster redshift. Figure \ref{overlay} shows the positional relationship
between the cluster galaxies -- as demarcated by red-sequence members --
and both the X-ray data and the strong lensing model. The peak of the
X-ray emission is coincident with the position of the BCG as is
typically seen in lower redshift relaxed clusters \citep{san09, bildfell08}. The center
of the overall distribution of the red sequence light is coincident
with the strong lensing mass peak, both of which are hence offset from
the BCG and the X-ray centroid by $\sim$60 kpc. Disagreements between
the mass peak as traced by lensing and the X-ray centroid are seen in
major clusters mergers \citep[e.g.][]{bra08,mah07,clo04} although the
magnitude of the disagreement in RCS2327 is not nearly as large and is
similar to that observed in intermediate cooling flow clusters in the
sample of clusters in \citet{all98}. However, the differing positions
indicated by various mass tracers is arguably the strongest evidence
that RCS2327 is anything but a single relaxed halo; we explore the
implications of this further in \S~\ref{s.mass}.

\section{Discussion}\label{s.mass}
The various mass proxies detailed in \S~\ref{s.followup} all indicate
that RCS2327 is an exceptionally
massive cluster given its redshift. Each of these mass proxies is naturally sensitive to
the mass of RCS2327 at a particular radius, and involves in all
cases one or more simplifying assumptions that allow the conversion of
the observable signal into a mass estimate. For example, the X-ray
data most directly constrain the mass at an overdensity radius of
$\sim$R$_{2500}$ and conversion of the X-ray spectrum and radial
luminosity profile to a spherical mass estimate requires the
assumption of hydrostatic equilibrium. The strong lensing data are
sensitive to mass at similar or smaller radii than the X-ray data, but
fundamentally measures a cylindrical mass in projection. The galaxy
dynamics are sensitive to mass at the virial scale and rely on
external scaling relations to provide a mass estimate, which, as
detailed in \S~\ref{s.spec}, are sensitive to not well known issues of velocity
bias and orbital anisotropies.

\subsection{Comparison to Other Clusters}\label{s.otherclusters}

\begin{figure*}
\centering
\includegraphics[scale=0.39]{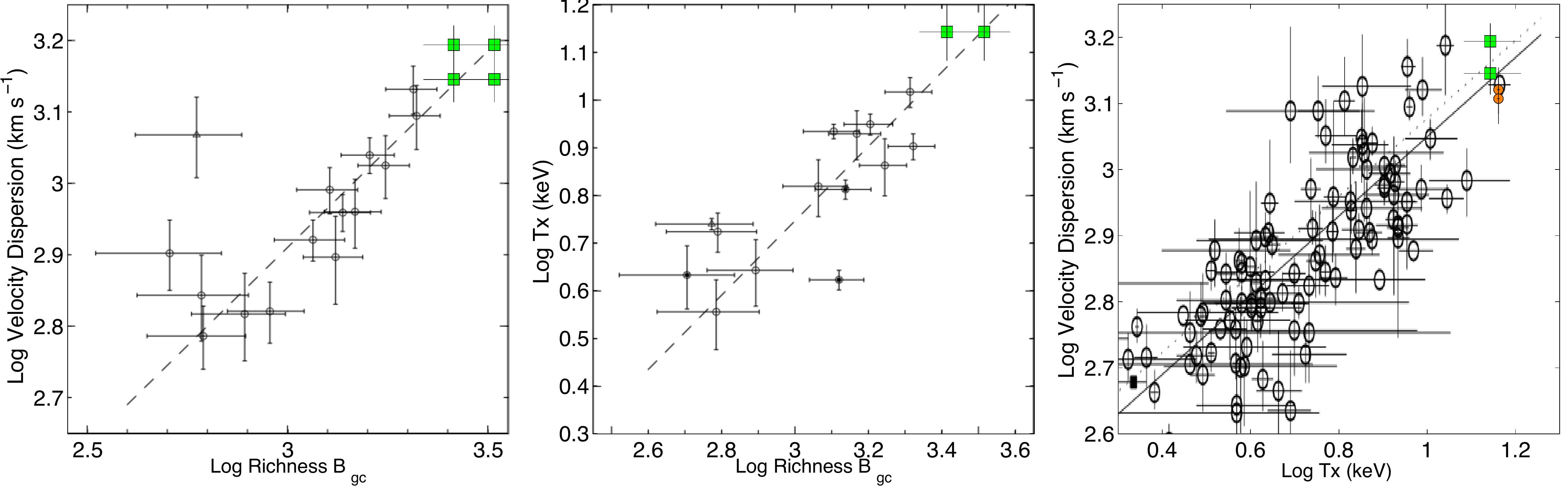} 
\caption{The comparison of
richness, velocity dispersion, and X-ray temperature for RCS2327
(green squares) against
relations for these quantities from the literature. The left and middle panels
compare to the data and fitted relations in \citet{yee03} for velocity
dispersion - richness (left) and X-ray temperature - richness (middle)
and the right panel compares to the data and fitted relation in
\cite{xue00} for velocity dispersion - X-ray temperature. Both the
total and early-type-only richnesses are shown, as well as velocity
dispersions from both the early-type-only and all galaxies. The X-ray
temperature and velocity dispersions (for all galaxies and early-type-only) of
\elgordo~ ($z=0.87$) from \citet{menanteau13} are plotted in
orange circles. 
}
\label{checkcorr}
\end{figure*}
The left and middle panels of Figure \ref{checkcorr} compare the velocity
dispersion, X-ray temperature and richness of RCS2327 to the global
correlations of these properties in an intermediate X-ray selected
cluster sample from \cite{yee03}. The right panel of Figure
\ref{checkcorr} plots the measured velocity dispersion and X-ray
temperature of RCS2327 against the cluster data and fitted relation
from \cite{xue00}. We plot both the total and red-sequence richnesses,
and the velocity dispersion from all galaxies, and only early-type
galaxies. Which of each of these properties is best compared to
the correlations in \cite{yee03} or \cite{xue00} is not obvious (e.g.,
see \S~\ref{s.richness}). Regardless, to within both the measurement uncertainties
and these systematic uncertainties, these three measures (which probe
large scale dynamics, the gas properties of the cluster core -- and
hence small scale dynamics, the gas fraction, and the like -- and the
stellar mass-to-light ratio) are all consistent with a massive cluster
with properties drawn from the global correlations seen in large
cluster samples. 

\begin{figure}
\centering
\includegraphics[scale=0.28]{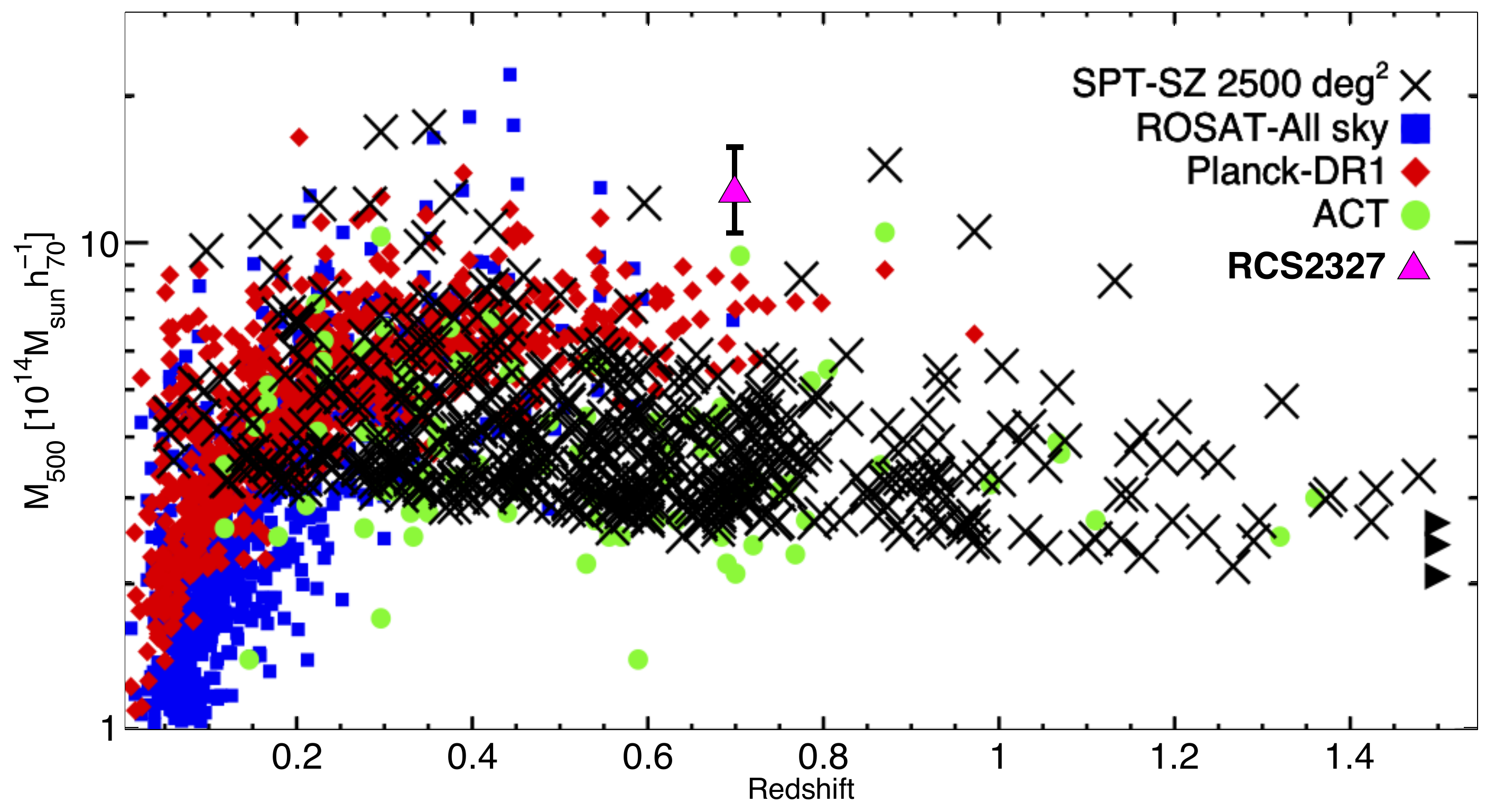} 
\caption{Reproduced from \citet{bleem14}. Estimated mass versus
  redshift for clusters from four large X-ray and SZ surveys: SPT-SZ 2500 deg$^2$
  \citep{bleem14}, {\it ROSAT} all sky survey \citep{piffaretti11},
  Planck-DR1 \citep{planck13}, and ACT \citep{marriage11b}. The
  X-ray mass of  
\uber~ ($M_{500}  = 1.3^{+ 0.3}_{-0.2} \times 10^{15}$\msun) is
  overplotted as a magenta triangle, placing it among the most massive
  clusters across all redshifts, and comparable to only few other
  clusters at $z\geq0.7$.   
We note that clusters that appear in different catalog may show
multiple times on this plot, e.g.,  \elgordo~ (\citet{marriage11b}, \citet{menanteau13}) appears with three mass estimates at
$z=0.87$. Other notable high-mass clusters are SPT-CLJ2337-5942 at
$z=0.77$, SPT-CLJ0615-5746 at $z=0.972$ and SPT-CLJ2106-5844 at
$z=1.132$  \citep{bleem14,foley11, planck11}. 
}
\label{bleem}
\end{figure}

In Figure~\ref{bleem}, reproduced from \citet{bleem14}, we plot the
estimated $M_{500}$ versus the redshift of \uber, compared to clusters
from large X-ray and SZ cluster surveys. The figure illustrates that
\uber~ is among the most massive clusters at all redshifts, and in
particular at $z\geq0.7$.

\subsection{Comparison of Mass Proxies}\label{s.massproxycomp}
Though each of the mass proxies discussed above measure the mass most
naturally at differing radii, it is still instructive to compare the
results directly. 
To do so we consider several additions
to the main analyses in \S~\ref{s.followup}. Table~\ref{allmass}
summarizes the mass estimates from different mass proxies at different
radii, and they are plotted in Figure~\ref{massfigure}.

\subsubsection{Cylindrical Masses from X-ray, Strong and Weak Lensing}\label{s.wlxray}
As noted above, weak and strong lensing are both sensitive to {\it projected} mass
density. However, they probe different regimes of the mass
distribution: strong lensing is insensitive to the mass at the
outskirts of the cluster, where no strong-lensing evidence
exists. Weak lensing lacks the resolution at the cluster core.  
To compare the weak and strong lensing mass estimates, 
we first compute the projected enclosed mass (also known as the aperture
mass) as a function of radius directly from the weak lensing data. 
{ We use the $\zeta_c$ statistic \citep{clo98,hoe07} and convert
the measurements into projected masses, using the best fit NFW to
estimate the large scale mean surface density \citep[see][for
details]{hoe07}. The dependence of the resulting projected mass estimate 
on the assumed density profile is minimal \citep{hoe15}. At a radius of 500 \h~ kpc this
yields a projected enclosed mass of $M_{\rm WL,cyl}(<500$\h$\rm{kpc})=5.7\pm1.1\times 10^{14}$\h\msun.
}

We can similarly extend the mass estimate from the strong lensing
model to larger radii. However, since the lens model is only
constrained by lensing evidence in the innermost { 400 kpc
(measured from the BCG) we increase the systematic uncertainty of the
strong lensing mass estimate by $\sim$10\%}.  
Following the analysis outlined in \S~\ref{s.richness}, we
find a mass at a 500\h kpc radius of $M_{\rm
SL,cyl}(<500$\h$\rm{kpc})=8.0\pm1.5 \times 10^{14}$\h\msun. 
 These two
values are in fair agreement. We refrain from extrapolating the strong
lensing mass to larger radii, where the strong lensing model is not
constrained.  

The X-ray masses can be converted to cylindrical mass, by
integrating along the line of sight out to 10 Mpc on both
sides of the cluster center. We note that this may introduce some uncertainty as
this is model-dependent.  
The projected enclosed X-ray masses at the radii of the lensed galaxies (see
Table~\ref{allmass}) are \MarcBX, \MarcAX, \MrfhX $ \times 10^{14}$\h\msun~ for 
271, 352, 500 kpc, respectively. These values are  
in fair agreement with the projected enclosed masses derived from
strong lensing,  \MarcB, \MarcA, \Mfhkpc $ \times 10^{14}$\h\msun,
respectively. The differences are in line with expected uncertainties
and biases (see, e.g., Mahdavi et al. 2013) for hydrostatic masses,
as overall we find that the lensing masses are somewhat higher than the
X-ray and SZ masses.
Nevertheless, it may also indicate that structure along the line
of sight or elongation of the cluster halo may be
significant.  For example, the structure that is indicated by a
concentration of galaxies at $z\sim0.73$ (Figure~\ref{dynrad}) may be
contributing to the lensing signal, and should be accounted for in future
lensing analysis \citep{daloisio13,mccully14,bayliss14}.

\subsubsection{Spherical Masses}\label{s.wlxray}
{ To compare the weak lensing, X-ray, and SZ masses we deproject the aperture masses following \cite{hoe07}, assuming the mass-concentration from \cite{duf08}. Although the deprojection is somewhat model dependent, it is less sensitive to deviations from
the NFW profile. At the cluster core, we compute
the corresponding deprojected weak lensing mass within
500 \h~kpc (approximately $R_{2500}$).
We obtain
a value of $M_{\rm WL}(<500$\h${\rm kpc})=4.1^{+1.2}_{-1.1}\times 10^{14}$\h\msun~ within this radius, in agreement with the
X-ray estimate of $M_{\rm{X},2500}=3.2^{+0.6}_{-0.3}\times
10^{14}$\h\msun, and SZ mass of $M_{\rm{SZ},2500}=4.6\pm0.5\times
10^{14}$\h\msun.}

At large radii,
we use the extrapolated X-ray mass as described in \S~\ref{s.xray}.
In making
this comparison we note that the native values of $R_{200}$ from each
of these analyses agree within the uncertainties. The X-ray mass is
$M_{\rm{X},200}=1.8^{+1.8}_{-0.7}\times 10^{15}$\h\msun~ and
the weak lensing mass from the NFW fit is
$M_{\rm{WL},200}(<2.1\rm{Mpc})=2.7\pm0.7\times 10^{15}$\h\msun.  
Hence at large radii the extrapolated X-ray mass and weak lensing
data also agree within the uncertainties.

\input{masstable2.tex}
\begin{figure*}
\centering
\includegraphics[scale=0.38]{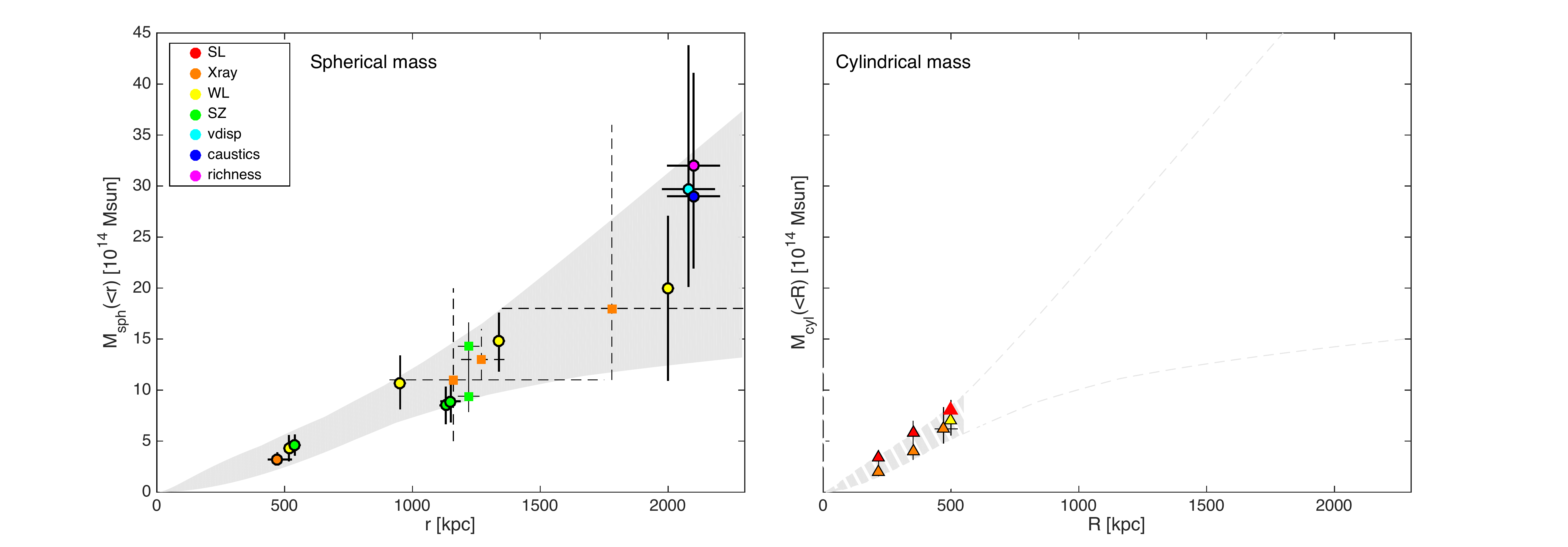} 
\caption{{ The mass estimates from the different mass proxies considered in
  this work (see Table~\ref{allmass}) are plotted as a function of
  radius, color-coded by mass proxy as indicated in the legend. 
  Points with dashed error bars are from extrapolated results (see text).   
The SZ mass estimates from \citet{hasselfield13} are plotted at
$R=1.2$ Mpc in green squares.
In the left panel we plot spherical masses within radius $r$, and in
the right panel are cylindrical (projected) masses enclosed within
  projected radius $R$. 1-$\sigma$ uncertainties are shown; we note
  that when $r_\Delta$ and $M_\Delta$ are determined jointly their
  uncertainties are correlated. 
The shaded area in the left panel is the 1-$\sigma$ range of spherical NFW mass profiles that were
fit to  the spherical masses measured in this work. The measurements
that were included in the fit are indicated with thick circles and errorbars. 
The
cylindrical masses in the right panel were not included in the fit,
nevertheless, we show the projected mass density of the same fits as
striped area in the right panel.  As
discussed in Section~\ref{s.massprofile}, this simple fit does not
represent a true joint analysis of the data, since various assumptions
on the slope of the
mass profile are already folded into some of these measurements.}}
\label{massfigure}
\end{figure*}

\subsection{Mass Profile}\label{s.massprofile}
Figure~\ref{massfigure} presents the enclosed masses measured in this
paper as a function of cluster-centric radius, as well as SZ masses
from the literature. As demonstrated above, these 
measurements are consistent with each other within errors, and trace 
the mass profile from the very core out to $R_{200}$. 

We fit a set of spherical NFW profiles (Eq.~\ref{eq.nfwmass}) to the
spherical masses measured in this paper. { To estimate the range of
  fits that are consistent with the measurements, we fit the profile
  1000 times, each time to a set of measurements that were randomly
sampled from their 1-$\sigma$ uncertainties, and weighted by their
  uncertainties. We did not include in the fit the extrapolated
  estimates and constraints from the literature. These masses are shown in
  Figure~\ref{massfigure} for reference, extrapolated measurements
  with in dashed error bars, and the Hasselfield et al. (2013)
  mass estimates in thin lines. }
A large range of scale radii is consistent with the measured masses,
and the resulting range of NFW profiles is shown as the solid shaded
area in Figure~\ref{massfigure}.
The striped area in the right panel of Figure~\ref{massfigure} traces
the cylindrical mass from the same NFW profiles that were fit to the
spherical masses.    While we could
simultaneously fit the profile to the cylindrical and spherical masses, we
choose not to, because the cylindrical strong lensing measurements do not assume spherical
symmetry and thus should not be expected to be described by a
spherical NFW profile. We find that the strong lensing masses are somewhat higher
than the predicted cylindrical masses, which could be due to the
triaxiality that is not taken into account in this simplified fit. As
expected, the projected X-ray masses do agree with the spherical profile,
since they were computed by integration of the X-ray best fit spherical
profile along the line of sight.  

{The simplistic NFW fit to all the non-extrapolated cylindrical
  mass measurements yields
  $r_s = 0.9_{-0.7}^{+1.3}{\rm Mpc}$}.
However, while a fit of a spherical NFW profile to the 
mass measurements is possible (though a large   
range of scale radii is consistent with the results), 
we caution that such a fit is not meaningful at this point.  
The different measurements were conducted completely independently
of each other, and rely on different assumptions as described in the
previous sections (e.g., mass-concentration relations, spherical
symmetry, hydrostatic equilibrium, various scaling
relations). In particular, some of the mass proxies already assume a certain
mass profile slope. 
A self-consistent combined multi-wavelength analysis  
is called for. Such an analysis would ideally allow triaxial
symmetry, and fit the mass distribution simultaneously 
to constraints derived directly from all the observables: strong lensing constraints, weak lensing
shear, galaxy velocity distribution, and X-ray and SZ measurements.
This sort of analysis is left for future work, and is not within the
scope of this paper. 

\section{Summary and Conclusions}\label{s.conclusions}
We present a multi-wavelength analysis of \clustername, a massive cluster at
$z=\clusterz$. The mass is estimated independently at several radii,
using seven different mass
proxies. At the core of the cluster, we measure the projected mass from
a strong lensing model; at intermediate radii, $\sim0.5$ Mpc, the mass
is estimated from X-ray, weak lensing, and the SZ effect. At large radii, $\sim2$
Mpc, we measure the cluster mass from its weak lensing signal, the
dynamics of galaxies in the cluster, and from scaling relations with
the richness of the cluster. This analysis provides a unique
opportunity of comparing methods and testing them against each other at a
significant redshift. 
In the previous section we compared mass estimates at overlapping
radii. Each of the mass proxies is prone to statistical and 
systematical uncertainties. Moreover, since all the measurements were
conducted independently from each other, some of the mass proxies rely on assumptions (e.g.,
assumed mass-concentration relation or the derived value of $r_\Delta$) that are
not necessarily uniform among these proxies.  
This unavoidably contributes to the scatter among the derived masses.
Nevertheless, the simple internal comparisons in \S~\ref{s.massproxycomp}, and
the comparison to global cluster correlations in \S~\ref{s.otherclusters} suggests that
RCS2327 is not a peculiar object (apart from its overall mass) and we
thus expect that a self consistent analysis would yield results
comparable to those presented here. 

In summary, all the evidence point to the conclusion that RCS2327 is
one of the most massive high redshift clusters known to date at
$z\geq0.7$. 

The set of measurements presented in this paper is expected to be
improved upon in the near future, with deep \hst~ 
observations that have already been executed. 
Further observations will provide constraints for a self-consistent
modeling of the three-dimensional cluster mass distribution
\citep[e.g.,][]{umetsu15, limousin13, sereno13},
that takes into account the effects of triaxiality and orientation on
the mass observables.

\acknowledgments
We wish to thank the anonymous referee for a constructive review that
improved the quality of this manuscript.
Support for program number GO-10846 was provided by NASA through a
grant from the Space Telescope Science Institute, which is operated by
the Association of Universities for Research in Astronomy, Inc., under
NASA contract NAS5-26555. 
Support for this work was provided by the National Aeronautics
and Space Administration through  {\it Chandra} award GO2-13158X issued by
the  {\it Chandra} X-ray Observatory Center, which is operated by the
Smithsonian Astrophysical Observatory for and on behalf of the
National Aeronautics Space Administration under contract NAS8-03060.
Based on observations obtained at the Gemini Observatory, which is operated by the 
Association of Universities for Research in Astronomy, Inc., under a cooperative agreement 
with the NSF on behalf of the Gemini partnership: the National Science Foundation 
(United States), the National Research Council (Canada), CONICYT (Chile), the Australian 
Research Council (Australia), Minist\'{e}rio da Ci\^{e}ncia, Tecnologia e Inova\c{c}\~{a}o 
(Brazil) and Ministerio de Ciencia, Tecnolog\'{i}a e Innovaci\'{o}n
Productiva (Argentina).
Based on observations obtained with MegaPrime/MegaCam, a joint project
of CFHT and CEA/DAPNIA, at the Canada-France-Hawaii Telescope (CFHT)
which is operated by the National Research Council (NRC) of Canada,
the Institute National des Sciences de l'Univers of the Centre
National de la Recherche Scientifique of France, and the University of
Hawaii. 
We also present observation taken at the Magellan telescopes at Las
Campanas Observatory, Chile, using LDSS-3 and GISMO. 
CARMA/SZA operations and science support is provided by the National
Science Foundation under a cooperative agreement and by the CARMA
partner universities; the CARMA/SZA work presented here was supported
by NSF grant AST- 1140019 to the University of Chicago.
ER acknowledges support from the National Science Foundation
AST-1210973, SAO TM3-14008X (issued under NASA Contract No. NAS8-
03060).  LFB research is funded by proyecto FONDECYT 1120676 and
Centro BASAL CATA.
ER acknowledges  support from FP7-PEOPLE-2013-IIF under Grant Agreement PIIF-GA-2013-627474.

{\it Facilities:} \facility{Magellan}, \facility{HST~ (ACS)},
\facility{CXO (ASIS)}, \facility{CFHT}, \facility{Gemini}.

\end{document}

%% file: masstable2.tex
\begin{deluxetable*}{l|lll|llllll}
\tablecolumns{10}
\tablewidth{0pc}
\tablecaption{ Estimated Masses \label{allmass}}
\tablehead{        
\colhead{Mass proxy}&
\multicolumn{3}{c}{Projected Mass [$10^{14}$\h\Msun] } &
\multicolumn{6}{c}{Spherical Mass $^\dagger$} \\
\colhead{} & 
\colhead{$R=217$kpc} & 
\colhead{$R=352$kpc} & 
\colhead{$R=500$kpc} &
\colhead{$r_{2500}$} & 
\colhead{$M_{2500}/10^{14}$} & 
\colhead{$r_{500}$} & 
\colhead{$M_{500}/10^{14}$} & 
\colhead{$r_{200}$} & 
\colhead{$M_{200}/10^{14} $} \\
\colhead{} & 
\colhead{} & 
\colhead{} & 
\colhead{} &
\colhead{[kpc]} & 
\colhead{ [\h\Msun]} & 
\colhead{[Mpc]} & 
\colhead{ [\h\Msun]} & 
\colhead{[Mpc]} & 
\colhead{ [\h\Msun]} 
}
\startdata
Strong Lensing & \MarcB  $\pm 0.5$&\MarcA $\pm 0.9$&[\Mfhkpc $\pm 1.4$]& \nodata &  \nodata & \nodata & \nodata & \nodata & \nodata \\
X-ray &\MarcBX & \MarcAX& \MrfhX  & $ 471^{+ 54}_{-33}$&$3.2^{+0.6}_{-0.3}$ & [$1.15^{+0.59}_{-0.25}$]&[$11^{+9}_{-6}$] & [$1.78^{+1.24}_{-0.43}$]&[$18^{+18}_{-7}$]\\
Weak Lensing & \nodata &  \nodata & $5.7\pm1.1$ & 517 & $4.3\pm1.2$ &1.34 & $15^{+2.9}_{-2.7}$  & \nodata&$20^{+9}_{-7}$\\
SZ & \nodata &  \nodata & \nodata  & \nodata & \nodata  & $1.13 \pm 0.02$&  $8.5\pm0.4$  & \nodata&\nodata\\
SZ  (M11$^{\ddagger}$) & \nodata &  \nodata & \nodata   & $540 \pm 6$& $4.6^{+0.1}_{-0.2}$  & $1.15\pm 0.02$&  $8.9\pm 0.8$  & \nodata&\nodata\\
Velocity dispersion& \nodata &  \nodata & \nodata  & \nodata& \nodata  & \nodata&  \nodata  & 2.1&$29.7^{+14}_{-9.5}$\\
Caustics & \nodata &  \nodata & \nodata & \nodata & \nodata & \nodata&  \nodata  & 2.1 &$29^{+10}_{-7}$\\
Richness & \nodata &  \nodata & \nodata & \nodata &  \nodata   & \nodata & \nodata& 2.1&$32^{+9}_{-8}$
\enddata
 \tablecomments{{ Summary of the mass estimates from the different mass proxies
   considered in this work. Square brackets indicate extrapolated values.} Projected X-ray mass was computed by
   integrating the mass model along the line of sight out to 10 Mpc on both
sides of the cluster.\\
$^\dagger$ { The different mass proxies were estimated within different
radii, as indicated.}\\$^{\ddagger}$ { SZ measurments using the
method of Mroczkowski (2011).}}
\end{deluxetable*}